\documentclass[aps,twocolumn,superscriptaddress]{revtex4}
\usepackage{graphicx}

\newcommand{\bra}[1] {\left\langle #1 \right|}
\newcommand{\ket}[1] {\left| #1 \right\rangle}

\begin{document}

\title{Weak and strong measurement of a qubit using a
switching-based detector}

\author{S. Ashhab}
\affiliation{Advanced Science Institute, The Institute of Physical
and Chemical Research (RIKEN), Wako-shi, Saitama 351-0198, Japan}
\affiliation{Physics Department, Michigan Center for Theoretical
Physics, The University of Michigan, Ann Arbor, Michigan
48109-1040, USA}

\author{J. Q. You}
\affiliation{Advanced Science Institute, The Institute of Physical
and Chemical Research (RIKEN), Wako-shi, Saitama 351-0198, Japan}
\affiliation{Department of Physics and Surface Physics Laboratory
(National Key Laboratory), Fudan University, Shanghai 200433,
China}

\author{Franco Nori}
\affiliation{Advanced Science Institute, The Institute of Physical
and Chemical Research (RIKEN), Wako-shi, Saitama 351-0198, Japan}
\affiliation{Physics Department, Michigan Center for Theoretical
Physics, The University of Michigan, Ann Arbor, Michigan
48109-1040, USA}

\date{\today}



\begin{abstract}
We analyze the operation of a switching-based detector that probes
a qubit's observable that does not commute with the qubit's
Hamiltonian, leading to a nontrivial interplay between the
measurement and free-qubit dynamics. In order to obtain analytic
results and develop intuitive understanding of the different
possible regimes of operation, we use a theoretical model where
the detector is a quantum two-level system that is constantly
monitored by a macroscopic system. We analyze how to interpret the
outcome of the measurement and how the state of the qubit evolves
while it is being measured. We find that the answers to the above
questions depend on the relation between the different parameters
in the problem. In addition to the traditional strong-measurement
regime, we identify a number of regimes associated with weak
qubit-detector coupling. An incoherent detector whose switching
time is measurable with high accuracy can provide high-fidelity
information, but the measurement basis is determined only upon
switching of the detector. An incoherent detector whose switching
time can be known only with low accuracy provides a measurement in
the qubit's energy eigenbasis with reduced measurement fidelity. A
coherent detector measures the qubit in its energy eigenbasis and,
under certain conditions, can provide high-fidelity information.
\end{abstract}

\maketitle

\section{Introduction}

The quantum theory of measurement is of particular significance in
the study of physics because it lies at the boundary between the
quantum and classical worlds \cite{Zurek}. Quantum measurement has
received renewed attention recently from a new perspective, namely
its role in quantum information processing \cite{Nielsen}.
Understanding how the quantum state evolves during the measurement
process and the information that can be extracted from the raw
measurement data is crucial for optimizing the design and
operation of measurement devices.

One typically thinks of the measurement process in terms of the
instantaneous, projective measurement encountered in the study of
basic quantum phenomena. For example, one sometimes imagines that
at any desired point in time the spin of a particle can be
measured along any desired axis. In many situations, however, the
measurement device is weakly coupled to the quantum system being
measured, such that the latter evolves almost freely according to
its free Hamiltonian while being barely perturbed by the measuring
device.

In the context of solid-state quantum information processing, the
weak coupling between the qubit and the measuring device is
imposed by the need to minimize the decoherence of the qubit
before the measurement; in practice it is not possible to
completely decouple the qubit from the measuring device, even
before the measurement is performed. It is worth mentioning here
that although this constraint of weak coupling might seem to be
undesirable for measurement purposes, weak coupling between the
qubit and the measuring device can have advantages over strong
coupling. For example, when the energy eigenstates of a
multi-qubit system contain some residual entanglement between the
qubits, this entanglement translates into errors when using strong
single-qubit measurements, whereas weak coupling to the measuring
device can lead to a measurement that is protected from such
errors \cite{Ashhab}.

In the study and implementation of qubit-state readout, one
strategy to remove any conflict between the qubit's free dynamics
and the measurement process has been to bias the qubit at a point
where the qubit Hamiltonian and the operator being probed by the
measuring device commute. In this case the free qubit dynamics
conserves the quantity being measured. This dynamics can therefore
be ignored during the measurement process, and the only difference
between weak and strong qubit-detector coupling is in the amount
of time needed to obtain a high-fidelity measurement result. It
should be noted, however, that in practice it is not always
possible to reach such an ideal bias point.

Here we are mainly interested in the analysis of the measurement
process when the qubit Hamiltonian and the operator being probed
by the measuring device do not commute. One must therefore analyze
the simultaneous working of two physical processes that, in some
sense, push the qubit in different directions. We shall focus on
the experimentally relevant example of a binary-outcome
switching-based measuring device (i.e.~the situation where the
measuring device can switch from its initial state, to which we
refer as 0, to another state, called 1, depending on the state of
the qubit), mainly as applied to superconducting circuits
\cite{YouReview}. Our work has a similar motivation to the recent
experimental and theoretical work of Refs.~\cite{Nakano,Tanaka}.
However, that work considered a measurement protocol where the
measured quantity is the value of the bias current at which
switching occurs, whereas we consider the more common approach
that uses the switching probability at fixed bias parameters.
Other related recent studies include work on the continuous
monitoring of Rabi oscillations \cite{Ilichev}, as well as the
theoretical study in Ref.~\cite{Wei} of multiple linear detectors
simultaneously probing different observables of the same qubit. A
models similar to ours was also used in Ref.~\cite{Brun} to study
certain aspects of quantum state evolution under the influence of
weak measurement.

It is worth clarifying from the outset a point concerning
terminology. A strong measurement is a measurement whose outcome
is strongly correlated with the measured quantity, thus specifying
with a high degree of certainty the state of the measured system.
After the measurement the system is, to a good approximation,
projected onto a state with a well-defined value of the measured
quantity. A weak measurement, on the other hand, is a measurement
whose outcome is weakly correlated with the measured quantity,
thus serving as a weak indicator of the state of the measured
system. Under suitable conditions a sequence composed of a large
number of weak measurements can result in an effectively strong
measurement. A weak measurement can also result in reliable
information when repeated a large number of times on identically
prepared systems. It is sometimes stated that weak measurement
results from weak coupling between the measured system and the
measuring device. It should be noted, however, that weak coupling
does not necessarily imply a weak measurement. We shall therefore
use the terms slow-switching and fast-switching detectors in this
paper. This distinction could be understood as being between
weak-coupling and strong-coupling measurements. A weak-coupling
measurement can result in a weak or strong measurement depending
on other system parameters, as we shall show with a number of
examples below.

This paper is organized as follows: In Sec.~II we describe the
model and briefly explain an example demonstrating the operation
of the switching-based detector in a simple case. In Sec.~III we
outline our approach to the analysis of the measurement process
when using a detector that has a short coherence time. A number of
specific examples are treated in Sec.~IV. The case of a detector
that has a relatively long coherence time is treated in Sec.~V.
The results of Secs.~II-V are reviewed in Sec.~VI. We discuss the
relevance of our analysis and results to the readout of
superconducting flux qubits in Sec.~VII, and we summarize our
results in Sec.~VIII.

\section{Model}

\begin{figure}[h]
\includegraphics[width=7.0cm]{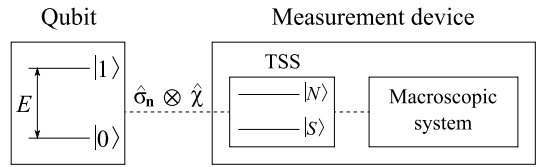}
\caption{Schematic diagram of the theoretical model that we use to
analyze the operation of a switching-based detector. The qubit's
energy eigenstates are denoted by $\ket{0}$ and $\ket{1}$, and
they have energy separation $E$. The detector is a two-state
system (TSS) [with a non-switched state $\ket{N}$ and a switched
state $\ket{S}$] that is constantly being monitored by a
macroscopic system. The detector probes the operator
$\hat{\sigma}_{\bf n}$ of the qubit. $\chi$ represents the
detector's operator through which the detector is coupled to the
qubit. The macroscopic system monitors whether the TSS remains in
the initially prepared state $\ket{N}$ or switches to the state
$\ket{S}$.}
\end{figure}

We consider the situation where one is trying to read out the
state of a quantum two-level system, i.e.~a qubit. The qubit is
placed in contact with a binary-outcome detector, i.e.~a detector
that produces one of two possible readings. The detector is
initialized in one of its two possible outcomes, and it can switch
to the other one, depending on the state of the qubit. For a large
part of the following analysis, we shall assume that the detector
itself is a quantum two-state system that is constantly being
monitored by a macroscopic, classical system (see Fig.~1). We
shall refer to the non-switched and switched states of the
detector as $\ket{N}$ and $\ket{S}$, respectively. Although the
two-state-detector model does not accurately describe the
realistic measuring devices that we shall discuss in Sec.~VII,
this simplification in the model will allow us to carry out our
calculations analytically and obtain a good amount of insight into
the dynamics of the measurement process. We shall come back to
this point and discuss the limitations of the simplified model in
Sec.~VII.

The detector is initially prepared in the state $\ket{N}$. We
shall assume that the switching is irreversible, such that after a
switching event the detector remains in the switched state until
it is reset by the experimenter. One possible way to naturally
prevent the detector from switching back to the state $\ket{N}$ is
to replace the state $\ket{S}$ by a continuum of states. However,
this change would results in lengthier algebra and similar results
to the two-state-detector model. We therefore do not make this
change, and we simply impose by hand the requirement that once the
detector is observed to be in the state $\ket{S}$ it does not
switch back to the state $\ket{N}$. For most of this paper, we
shall also ignore any intrinsic qubit decoherence (part of which
could be caused by the presence of the detector) that is not
required by a correct quantum description of the measurement
process.

It is probably most conventional to describe the qubit using a
basis where the detector probes the operator $\hat{\sigma}_z$ of
the qubit, where $\hat{\sigma}_{x,y,z}$ are the Pauli matrices. In
other words, the detector probes the $z$ component of the
pseudospin associated with the qubit. The qubit Hamiltonian then
points along an arbitrary direction. For the analysis below, we
find it easier to describe the qubit in the energy eigenbasis. The
qubit Hamiltonian is therefore expressed as
\begin{equation}
\hat{H}_{\rm q} = \left( \begin{array}{cc}
-\frac{E}{2} & 0 \\
0 & \frac{E}{2} \\
\end{array}
\right) = - \frac{E}{2} \hat{\sigma}_z,
\end{equation}
where $E$ is the energy splitting between the qubit's two energy
levels. We shall express the ground and excited states of the
Hamiltonian as $\ket{0}$ and $\ket{1}$, respectively. The qubit
operator that the detector probes (i.e.~the qubit property that
determines the two different switching rates of the detector) can
be expressed as $\hat{\sigma}_{\bf n} = (\cos\beta) \hat{\sigma}_z
+ (\sin\beta) \hat{\sigma}_x$. We shall express the eigenstates of
this operator as
\begin{eqnarray}
\ket{L} & = & \cos\frac{\beta}{2} \ket{0} + \sin\frac{\beta}{2}
\ket{1} \nonumber
\\
\ket{R} & = & \sin\frac{\beta}{2} \ket{0} - \cos\frac{\beta}{2}
\ket{1}.
\label{Eq:Two_bases}
\end{eqnarray}

We postpone writing down a Hamiltonian of the system because, as
we shall see below, the results of Secs.~III and IV are
independent of the details of the Hamiltonian and the analysis can
be performed without knowledge of these details, while the
behaviour of the detector in the case analyzed in Sec.~V depends
on these details. Therefore, by writing down a Hamiltonian at this
point we run the risk of losing some insight into how the
measurement process proceeds. Some specific Hamiltonians will be
given in Sec.~V.

\subsection{Simple case}

Let us consider the following simple case for demonstration
purposes: We take $\beta=0$, such that $\ket{L}=\ket{0}$ and
$\ket{R}=\ket{1}$, up to an irrelevant phase factor. In this case
we do not need to worry about any mixing dynamics between the
states $\ket{L}$ and $\ket{R}$ during the measurement process. If
we now take the switching rate for one of the two states (say
$\ket{L}$) to be exactly zero and the switching rate for the other
state (here $\ket{R}$) to be $\gamma$, we find that one can
perform the measurement by connecting the qubit to the detector
and waiting a duration $\tau$ that is long compared to $1/\gamma$.
If the detector switches to the state $\ket{S}$, we know for sure
that the qubit was in the state $\ket{R}$. If the detector remains
in the state $\ket{N}$, we can say with a high degree of certainty
that the qubit was in the state $\ket{L}$. In other words,
instances where $\gamma\tau\gg 1$ and the qubit is in the state
$\ket{R}$ but the detector does not switch should be very rare and
can be ignored for all practical purposes. We can therefore reach
essentially 100\% measurement fidelity using this procedure.

The main reason why understanding this case was simple is that we
did not need to worry about any mixing dynamics between the states
$\ket{L}$ and $\ket{R}$ during the measurement process. This
simple picture breaks down when $\beta\neq 0$. In that case, one
could say that the Hamiltonian tries to push the qubit's state to
precess around a certain axis (determined by the qubit's
Hamiltonian), while the detector tries to slowly pull the qubit's
state towards a different axis (determined by the probed
operator). The interplay between these two mechanisms gives rise
to nontrivial dynamics and measurement scenarios. The analysis of
this interplay is the main subject of this paper and will be
carried out next.

\section{Incoherent detector}

We start with the case where the coherence time of the detector
(i.e. the dephasing time for coherent superpositions of the states
$\ket{N}$ and $\ket{S}$) is short compared to the inverse of the
qubit's energy scale $E$ (with the unit-conversion factor
$\hbar$). This limit could also be seen as the limit of a
classical detector. Note that with the above constraint we are not
imposing any requirement on the relation between the detector's
switching rate and the energy scale $E$.

\subsection{Qubit dynamics during a short period of time}

In order to develop a theoretical understanding of the qubit
dynamics during the measurement process, we consider a period of
time $\delta t$ that is short enough that (1) the detector's
switching probability during this period is small and (2) the
qubit Hamiltonian causes a negligible change in the qubit's state,
but the same period of time is long enough that the switching
probability scales linearly with time (for time intervals that are
short compared to the coherence time of the detector, the
detector's dynamics resembles coherent oscillations between the
states $\ket{N}$ and $\ket{S}$; for time intervals that are long
compared to the coherence time of the detector, the detector
exhibits exponential-decay-like dynamics from the state $\ket{N}$
to the state $\ket{S}$). The fact that the detector has a very
short coherence time is crucial for the simultaneous
satisfiability of the above conditions. Note that, apart from its
role in deriving some differential equations below, $\delta t$
does not have any particular significance. It can therefore be
chosen freely within the range specified above.

We identify the short time interval described above by its initial
and final times, $t$ and $t+ \delta t$. If during this interval
the qubit is in the state $\ket{L}$, the detector will switch with
rate $\gamma_L$, resulting in a switching probability $\gamma_L
\delta t$. If, on the other hand, the qubit is in the state
$\ket{R}$, the detector will switch with probability $\gamma_R
\delta t$.

The evolution of the qubit's state can be described using the
density matrix $\rho_{\rm q}(t)$ at time $t$:
\begin{equation}
\rho_{\rm q}(t) = \left(
\begin{array}{cc}
\rho_{00}(t) & \rho_{01}(t) \\
\rho_{10}(t) & \rho_{11}(t) \\
\end{array}
\right).
\end{equation}
We can allow the density matrix to have a trace smaller than one
(this `unnormalized' trace would correspond to the probability
that the detector has not switched from the beginning of the
measurement until time $t$). As mentioned above, we assume that
the detector can switch for both states $\ket{L}$ and $\ket{R}$,
with rates $\gamma_L$ and $\gamma_R$ (we shall generally take
$\gamma_R>\gamma_L$).

Assuming that the detector has not switched until time $t$, the
state of the entire system composed of the qubit and the detector
is given by the simple product
\begin{equation}
\rho_{\rm Total} (t) = \rho_{\rm q}(t) \otimes \ket{N}\bra{N}.
\end{equation}
Because of the qubit-detector interaction, the Hamiltonian takes
the density matrix of the entire system into the state
\begin{equation}
\rho_{\rm Total} (t+\tilde{\delta t}) = \hat{U}_{\rm Total}
(\tilde{\delta t}) \rho_{\rm Total} (t) \hat{U}_{\rm
Total}^{\dagger} (\tilde{\delta t})
\label{Eq:Evolution_total_system_coherent}
\end{equation}
at time $t+\tilde{\delta t}$, where $\tilde{\delta t}$ is taken to
be so short that all dynamics is coherent (the superscript
$\dagger$ indicates the transpose conjugate of a matrix; note that
several matrices in this paper will not be unitary, even though
some of them will be denoted using the letter $U$). Since the time
interval under consideration, i.e.~$t$ to $t+\delta t$, is long
enough that the switching follows exponential-decay-like dynamics,
we have to modify Eq.~(\ref{Eq:Evolution_total_system_coherent})
such that no coherence between the states $\ket{N}$ and $\ket{S}$
appears on that timescale. Making the proper adjustments, and
neglecting for a moment the qubit's free evolution and other phase
shifts that could be caused by the measurement but can be
straightforwardly accounted for, we find that the evolution of the
total density matrix is governed by the formula
\begin{eqnarray}
\rho_{\rm Total} (t+\delta t) & = & \hat{U}_{\rm Total,s} (\delta
t) \rho_{\rm Total} (t) \hat{U}_{\rm Total,s}^{\dagger} (\delta t)
\nonumber \\ & & + \hat{U}_{\rm Total,ns} (\delta t) \rho_{\rm
Total} (t) \hat{U}_{\rm Total,ns}^{\dagger} (\delta t),
\label{Eq:Evolution_total_system_incoherent}
\end{eqnarray}
where
\begin{widetext}
\begin{eqnarray}
\hat{U}_{\rm Total,s} (\delta t) & = & \sqrt{\gamma_L \delta t}
\ket{L}\bra{L} \otimes \ket{S}\bra{N} + \sqrt{\gamma_R \delta t}
\ket{R}\bra{R} \otimes \ket{S}\bra{N} \nonumber
\\
\hat{U}_{\rm Total,ns} (\delta t) & = & \sqrt{1-\gamma_L \delta t}
\ket{L}\bra{L} \otimes \ket{N}\bra{N} + \sqrt{1-\gamma_R \delta t}
\ket{R}\bra{R} \otimes \ket{N}\bra{N}.
\end{eqnarray}
One can now imagine that a measurement is performed on the
detector at time $t+\delta t$, projecting it either onto the
non-switched state $\ket{N}$ or the switched state $\ket{S}$. The
state of the qubit evolves accordingly. The qubit's new density
matrix can be expressed as
\begin{equation}
\rho(t+\delta t) = \hspace{0.3cm} \hat{P}_{\rm s}(\delta t)
\rho(t) \hat{P}_{\rm s}^{\dagger}(\delta t) \hspace{0.3cm} {\rm
or} \hspace{0.3cm} \hat{P}_{\rm ns}(\delta t) \rho(t) \hat{P}_{\rm
ns}^{\dagger}(\delta t),
\end{equation}
depending on the measured state of the detector (note that we have
dropped the index q identifying the qubit's density matrix; from
now on we only consider the evolution of the qubit's density
matrix). The operators $\hat{P}_{\rm s}(\delta t)$ and
$\hat{P}_{\rm ns}(\delta t)$ can be derived straightforwardly from
Eq.~(\ref{Eq:Evolution_total_system_incoherent}), and we shall
give expressions for them shortly.

To summarize the above argument, the qubit's density matrix
evolves in one of two possible ways during the interval $t$ to $t
+ \delta t$, depending on whether the detector switches or not.
The measurement-induced part of the evolution is described by the
operators $\hat{P}_{\rm s}(\delta t)$ and $\hat{P}_{\rm ns}(\delta
t)$, which correspond to switching and no-switching events during
the interval of length $\delta t$, respectively. These projection
operators are given by
\begin{eqnarray}
\hat{P}_{\rm s}(\delta t) & = & \sqrt{\gamma_L \delta t} \ket{L}
\bra{L} + \sqrt{\gamma_R \delta t} \ket{R} \bra{R} \nonumber \\
& = & \sqrt{\gamma_L \delta t} \left(
\begin{array}{cc}
\cos^2\frac{\beta}{2} & \sin\frac{\beta}{2}\cos\frac{\beta}{2} \\
\sin\frac{\beta}{2}\cos\frac{\beta}{2} & \sin^2\frac{\beta}{2} \\
\end{array}
\right)+ \sqrt{\gamma_R \delta t} \left( \begin{array}{cc}
\sin^2\frac{\beta}{2} & -\sin\frac{\beta}{2}\cos\frac{\beta}{2} \\
-\sin\frac{\beta}{2}\cos\frac{\beta}{2} & \cos^2\frac{\beta}{2} \\
\end{array}
\right) \nonumber
\\
\hat{P}_{\rm ns}(\delta t) & = & \sqrt{1-\gamma_L \delta t}
\ket{L} \bra{L} + \sqrt{1-\gamma_R \delta t} \ket{R} \bra{R} \nonumber \\
& \approx & \left(
\begin{array}{cc}
1 - \frac{\gamma_+ \delta t}{2} + \frac{\gamma_- \delta t}{2}
\cos\beta
& \frac{\gamma_- \delta t}{2}\sin\beta \\
\frac{\gamma_- \delta t}{2}\sin\beta & 1 - \frac{\gamma_+ \delta
t}{2} - \frac{\gamma_- \delta t}{2}\cos\beta \\
\end{array}
\right),
\label{Eq:Basic_projection_operators}
\end{eqnarray}
\end{widetext}
where $\gamma_{\pm} = \left( \gamma_R \pm \gamma_L \right)/2$.

The probability that the detector will switch during the time
interval $t$ to $t + \delta t$ is given by
\begin{eqnarray}
{\rm Prob}_{\rm s}(t,t+\delta t) & = & {\rm Tr} \left\{
\hat{P}_{\rm s}(\delta t) \rho(t) \hat{P}_{\rm s}^{\dagger}(\delta
t) \right\}.
\end{eqnarray}
If the detector switches, the qubit's density matrix immediately
after the switching event will be given by
\begin{equation}
\hat{\rho}(t+\delta t) \approx \hat{P}_{\rm s}(\delta t) \rho(t)
\hat{P}_{\rm s}^{\dagger}(\delta t),
\label{Eq:Evolved_rho_switch}
\end{equation}
up to the usual renormalization constant that is not needed for
our purposes. The transformation given in
Eq.~(\ref{Eq:Evolved_rho_switch}) describes a substantial change
in the state of the qubit. Therefore, the small change induced by
the qubit's Hamiltonian during this short interval can be ignored.
Note that in this paper we shall not be interested in any coherent
dynamics that might occur after a switching event. As a result, it
is not important for our purposes that coherence between the
states $\ket{L}$ and $\ket{R}$ be maintained after a switching
event.

If the detector does not switch between $t$ and $t+\delta t$, the
qubit's density matrix will be transformed according to the
operator $\hat{P}_{\rm ns}(\delta t)$. Since the change induced by
$\hat{P}_{\rm ns}(\delta t)$ is small (proportional to $\delta
t$), one must now take into account the comparably small change
induced by the qubit Hamiltonian, which is described by the
unitary transformation
\begin{equation}
\hat{U}_{\rm Ham}(\delta t) = e^{-i\hat{H} \delta t} = \left(
\begin{array}{cc}
e^{iE\delta t/2} & 0 \\
0 & e^{-iE\delta t/2} \\
\end{array}
\right).
\label{Eq:Free_evolution_operator}
\end{equation}
The qubit's density matrix at time $t+\delta t$ will therefore be
given by
\begin{equation}
\rho(t+\delta t) \approx \hat{U}_{\rm Ham}(\delta t) \hat{P}_{\rm
ns}(\delta t) \rho(t) \hat{P}_{\rm ns}^{\dagger}(\delta t)
\hat{U}_{\rm Ham}^{\dagger}(\delta t).
\label{Eq:Evolved_rho_1}
\end{equation}
Note that $\hat{U}_{\rm Ham}(\delta t)$ and $\hat{P}_{\rm
ns}(\delta t)$ can be treated as commuting operators in the limit
$\delta t \rightarrow 0$; as mentioned above, we are assuming that
$\delta t$ is much smaller than both $1/E$ and $1/\gamma_+$.

It is worth pausing for a moment here to comment on an issue
related to the terminology of strong versus weak measurement.
There are two possible outcomes of the measurement during the
interval $t$ to $t+\delta t$. If the detector switches, the qubit
undergoes a large change (Eq.~\ref{Eq:Evolved_rho_switch}). In
particular, if $\gamma_L=0$, the measurement would constitute a
strong measurement. A no-switching event, on the other hand,
causes a very small change in the state of the qubit
(Eq.~\ref{Eq:Evolved_rho_1}). Whether such a measurement is weak
or strong therefore depends on the outcome of the measurement. One
could say that since the strong-measurement outcome occurs with a
very small probability, the measurement performed during the
interval of length $\delta t$ is a weak-measurement on average.

\subsection{Equation of motion for the evolution operator before
switching}

We now take the density matrix of a qubit in an experimental run
where no switching has occurred (yet) and express it as
\begin{equation}
\rho(t) = \hat{U}_{\rm ns}(t) \rho(0) \hat{U}_{\rm
ns}^{\dagger}(t).
\label{Eq:Evolved_rho_2}
\end{equation}
The operator $\hat{U}_{\rm ns}(t)$ is therefore the propagator
that describes how the state of the qubit evolves from the initial
time $t=0$ until time $t$. Comparing Eqs.~(\ref{Eq:Evolved_rho_1})
and (\ref{Eq:Evolved_rho_2}), we find that
\begin{eqnarray}
\hat{U}_{\rm ns}(t+\delta t) & \approx & \hat{U}_{\rm Ham}(\delta
t) \hat{P}_{\rm ns}(\delta t) \hat{U}_{\rm ns}(t) \nonumber \\
& = & \hat{U}_{\rm ns}(t) + \left[ \hat{U}_{\rm Ham}(\delta t)
\hat{P}_{\rm ns}(\delta t) - 1 \right] \hat{U}_{\rm ns}(t).
\nonumber \\
\end{eqnarray}
Using Eqs.~(\ref{Eq:Basic_projection_operators}) and
(\ref{Eq:Free_evolution_operator}) and taking the limit $\delta t
\rightarrow 0$, we find that the operator $\hat{U}_{\rm ns}(t)$
obeys the differential equation
\begin{widetext}
\begin{equation}
\frac{d \hat{U}_{\rm ns}(t)}{dt} = \left(
\begin{array}{cc}
i \frac{E}{2} - \frac{\gamma_+}{2} + \frac{\gamma_-}{2}\cos\beta
& \frac{\gamma_-}{2}\sin\beta \\
\frac{\gamma_-}{2}\sin\beta & - i \frac{E}{2} - \frac{\gamma_+}{2}
-
\frac{\gamma_-}{2}\cos\beta \\
\end{array}
\right) \hat{U}_{\rm ns}(t).
\end{equation}
This linear differential equation can be solved straightforwardly
to give
\begin{equation}
\hat{U}_{\rm ns} (t) = \left(
\begin{array}{ccc}
\cos^2\frac{\eta}{2} e^{\lambda_+ t} + \sin^2\frac{\eta}{2}
e^{\lambda_- t} & & \sin\frac{\eta}{2}\cos\frac{\eta}{2}
\left( e^{\lambda_+ t} - e^{\lambda_- t} \right) \\
\\
\sin\frac{\eta}{2}\cos\frac{\eta}{2} \left( e^{\lambda_+ t} -
e^{\lambda_- t} \right) & & \sin^2\frac{\eta}{2} e^{\lambda_+ t} +
\cos^2\frac{\eta}{2} e^{\lambda_- t}
\end{array}
\right),
\label{Eq:U_no_switch}
\end{equation}
\end{widetext}
where
\begin{eqnarray}
\lambda_{\pm} & = & - \frac{\gamma_+}{2} \pm \frac{1}{2}
\sqrt{\left( \gamma_-^2 - E^2 \right) + 2 i E \gamma_- \cos\beta}
\nonumber
\\
\tan\eta & = & \frac{\gamma_- \sin\beta}{\gamma_- \cos\beta + i
E}.
\label{Eq:lambda_and_phi}
\end{eqnarray}

Similarly, the qubit's density matrix after a switching event that
occurs between $t$ and $t + \Delta t$ ($\Delta t$ can be
understood as the accuracy with which the switching time can be
determined, and for now it is assumed to be much smaller than
$1/E$) can be expressed as
\begin{equation}
\rho(t,\Delta t) = \hat{U}_{\rm s}(t,\Delta t) \rho(0)
\hat{U}_{\rm s}^{\dagger}(t,\Delta t),
\end{equation}
where
\begin{equation}
\hat{U}_{\rm s}(t,\Delta t) = \hat{P}_{\rm s}(\Delta t)
\hat{U}_{\rm ns}(t).
\label{Eq:U_switch}
\end{equation}

The above expressions will form the basis for the analysis of the
remainder of this section and that of Sec.~IV. They describe, in a
general setting, how the qubit's state evolves during the
measurement process. As we shall see below, they also describe
what information can be extracted from a given measurement
outcome.

\subsection{Measurement basis and fidelity for a given outcome}

There are a large number of possible outcomes of a measurement
attempt. One possibility is that the detector does not switch
throughout a measurement of duration $\tau$. In this case the
corresponding evolution operator that gives the qubit's density
matrix at time $\tau$ is simply given by $\hat{U}_{\rm ns} (\tau)$
from Eq.~(\ref{Eq:U_no_switch}). All the other possibilities are
described by instances where the detector does not switch during
the interval between the initial time ($t=0$) and time $t$ and
switches between times $t$ and $t+\Delta t$. In this case, the
relevant evolution operator is $\hat{U}_{\rm s} (t,\Delta t)$.

The question now is what information (about the initial state of
the qubit) one can extract from a given outcome, e.g.~a detector
switching event that occurs between times $t$ and $t+\Delta t$ or
a no-switching instance after time $\tau$. In other words, what
would be the measurement basis and fidelity that correspond to a
given experimental outcome? It should be emphasized here that, in
general, the measurement basis and fidelity depend not only on the
measurement procedure, but also on the specific outcome obtained
in a given experimental run.

Once a given outcome is observed, one can take the relevant
evolution matrix $\hat{U}_j$, which will be one of the matrices
$\hat{U}_{\rm s} (t,\Delta t)$ and $\hat{U}_{\rm ns} (t)$, and
think of it as being composed of two parts: a measurement operator
followed by a rotation,
\begin{equation}
\hat{U}_{j} = \hat{U}_{j}^{\rm rot} \hat{U}_{j}^{\rm meas}.
\end{equation}
The measurement operator takes the form
\begin{equation}
\hat{U}_{j}^{\rm meas} = \sqrt{P_{j}^{(1)}} \ket{\psi_{j}^{(1)}}
\bra{\psi_{j}^{(1)}} + \sqrt{P_{j}^{(2)}} \ket{\psi_{j}^{(2)}}
\bra{\psi_{j}^{(2)}},
\label{Eq:U_meas}
\end{equation}
where $\ket{\psi_{j}^{(1)}}$ and $\ket{\psi_{j}^{(2)}}$ are two
orthogonal states that represent the measurement basis of this
particular outcome, and $P_{j}^{(1)}$ and $P_{j}^{(2)}$ are,
respectively, the probabilities that the specific outcome under
consideration will be observed if the qubit was initially in state
$\ket{\psi_{j}^{(1)}}$ or $\ket{\psi_{j}^{(2)}}$.

For a given initial state $\rho(0)$, the probability that the
specific outcome under consideration occurs in an experimental run
is given by
\begin{eqnarray}
{\rm Prob}_{j} & = & {\rm Tr} \left\{ \hat{U}_{j}
\rho(0) \hat{U}_{j}^{\dagger} \right\} \nonumber \\
& = & {\rm Tr} \left\{ \hat{U}_{j}^{\dagger}
\hat{U}_{j} \rho(0) \right\} \nonumber \\
& = & {\rm Tr} \left\{ \left( \hat{U}_{j}^{\rm meas} \right)^2
\rho(0) \right\}.
\end{eqnarray}
As a result, by diagonalizing the measurement operator
$\hat{U}_{j}^{\rm meas}$ (or equivalently, by diagonalizing
$\hat{U}_{j}^{\dagger} \hat{U}_{j}$), we can obtain the
measurement basis and fidelity of the performed measurement.

The measurement basis is defined by the states
$\ket{\psi_{j}^{(1)}}$ and $\ket{\psi_{j}^{(2)}}$. The measurement
fidelity is given by the difference between the probability of
making a correct inference about the state of the qubit and the
probability of making a wrong inference about the state of the
qubit. In order to describe all possible input states, one must
calculate the fidelity by taking a statistical average using the
maximally mixed state
\begin{equation}
\rho_{\rm max. \ mixed} = \left( \begin{array}{cc}
\frac{1}{2} & 0 \\
0 & \frac{1}{2} \\
\end{array} \right).
\label{Eq:Maximally_mixed}
\end{equation}
If we choose the labels in Eq.~(\ref{Eq:U_meas}) such that
$P_j^{(1)}>P_j^{(2)}$, one would maximize the probability of
making a correct inference about the qubit's state by associating
the specific outcome $j$ with the state $\ket{\psi_j^{(1)}}$.
Using the maximally mixed state, a straightforward calculation
shows that the fidelity in this case is given by
\begin{equation}
F_j = \frac{P_{j}^{(1)} - P_{j}^{(2)}}{P_{j}^{(1)} + P_{j}^{(2)}}.
\end{equation}

When there are more than one possible outcome for the measuring
device (which is always the case for any useful measuring device),
the overall fidelity can be evaluated by taking the statistical
average of the fidelities for the different possible outcomes.
This averaging procedure will be performed in Sec.~IV. We should
emphasize, however, that the fidelities of the different specific
outcomes also carry some significance of their own. This point
will become clearer when we discuss a number of illustrative
examples below.

An alternative approach to deriving the fidelity goes as follows:
let us assume that we initially have no information about the
state of the qubit. The state is therefore described by the
maximally mixed state in Eq.~(\ref{Eq:Maximally_mixed}). After the
detector gives the outcome denoted by the index $j$, we have
partial or complete knowledge about the qubit's state immediately
after the measurement, or in other words the collapsed qubit
state. The amount of information that we have gained can be
quantified by the purity of the qubit's state after the
measurement;
\begin{equation}
{\rm Purity} = \sqrt{2 {\rm Tr} \left\{ \left( \hat{U}_{j}
\rho_{\rm max. \ mixed} \hat{U}_{j}^{\dagger} \right)^2 \right\}
-1}.
\end{equation}
A straightforward calculation shows that this expression agrees
with the one derived above for the measurement fidelity.

\section{Analyzing different important cases}

The general expressions that we have derived in Sec.~III do not
provide a clear, intuitive picture of how the qubit's state
evolves during the measurement and how the measurement data should
be interpreted. We therefore analyze some special, representative
cases below. Section IV.A analyzes the simple case discussed in
Sec.~II.A, while the choice of the cases analyzed in Secs. IV.B
and IV.C is prompted by the fact that
Eq.~(\ref{Eq:lambda_and_phi}) can be simplified in these opposite
limits.

\subsection{Case 1: $\beta=0$ (No mixing between the states
$\ket{L}$ and $\ket{R}$ during the measurement)}

The case $\beta=0$ is the case where the Hamiltonian and
measurement axes are parallel (in other words, the states
$\ket{L}$ and $\ket{R}$ coincide with the states $\ket{0}$ and
$\ket{1}$). We can therefore expect to reproduce the simple, known
results outlined in Sec.~II.A. When $\beta=0$ (which also gives
$\eta=0$), we find that
\begin{eqnarray}
\hat{P}_{\rm s}(\Delta t) & = & \left(
\begin{array}{cc}
\sqrt{\gamma_L \Delta t} & 0 \\
0 & \sqrt{\gamma_R \Delta t} \\
\end{array}
\right) \nonumber
\\
\hat{U}_{\rm ns} (t) & = & \left(
\begin{array}{cc}
e^{\lambda_+ t} & 0 \\
0 & e^{\lambda_- t}
\end{array}
\right),
\end{eqnarray}
with
\begin{equation}
\lambda_{\pm} = -\frac{\gamma_+}{2} \pm \frac{\gamma_- + i E}{2}.
\end{equation}
We therefore find that
\begin{eqnarray}
\hat{U}_{\rm ns} (t) & = & \left(
\begin{array}{cc}
e^{i \frac{E}{2} t} & 0 \\
0 & e^{- i \frac{E}{2} t}
\end{array}
\right) \hat{U}_{\rm ns}^{\rm meas} (t) \nonumber
\\
\hat{U}_{\rm s} (t,\Delta t) & = & \left(
\begin{array}{cc}
e^{i \frac{E}{2} t} & 0 \\
0 & e^{- i \frac{E}{2} t}
\end{array}
\right) \hat{U}_{\rm s}^{\rm meas} (t,\Delta t),
\end{eqnarray}
where
\begin{eqnarray}
\hat{U}_{\rm ns}^{\rm meas} (t) & = & \left(
\begin{array}{cc}
e^{- \frac{\gamma_L}{2} t} & 0 \\
0 & e^{- \frac{\gamma_R}{2} t}
\end{array}
\right) \nonumber
\\
\hat{U}_{\rm s}^{\rm meas} (t,\Delta t) & = & \left(
\begin{array}{cc}
\sqrt{\gamma_L \Delta t} e^{- \frac{\gamma_L}{2} t} & 0 \\
0 & \sqrt{\gamma_R \Delta t} e^{- \frac{\gamma_R}{2} t} \\
\end{array}
\right).
\label{Eq:Umeas_zero_beta}
\end{eqnarray}
The eigenvectors of the matrices $\hat{U}_{\rm s}^{\rm meas}
(t,\Delta t)$ and $\hat{U}_{\rm ns}^{\rm meas} (t)$ are the states
$\ket{L}$ and $\ket{R}$ (independently of the values of $t$ and
$\Delta t$). The measurement basis is therefore $\{ |L\rangle ,
|R\rangle \}$, as expected. The eigenvalues of $\hat{U}_{\rm
s}^{\rm meas} (t,\Delta t)$ are given by
\begin{eqnarray}
\sqrt{P_{\rm s}^{L}(t,t+\Delta t)} & = & \sqrt{\gamma_L \Delta t}
\exp \left\{ - \frac{\gamma_L t}{2} \right\} \nonumber
\\
\sqrt{P_{\rm s}^{R}(t,t+\Delta t)} & = & \sqrt{\gamma_R \Delta t}
\exp \left\{ - \frac{\gamma_R t}{2} \right\},
\label{Eq:Case1_Eigenvalues}
\end{eqnarray}
and those of $\hat{U}_{\rm ns}^{\rm meas} (t)$ are given by
\begin{eqnarray}
\sqrt{P_{\rm ns}^{L}(t)} & = & \exp \left\{ - \frac{\gamma_L t}{2}
\right\} \nonumber
\\
\sqrt{P_{\rm ns}^{R}(t)} & = & \exp \left\{ - \frac{\gamma_R t}{2}
\right\}.
\end{eqnarray}

\subsubsection{Fidelity of a specific outcome}

The eigenvalues in Eq.~(\ref{Eq:Case1_Eigenvalues}) give a
fidelity (for switching time $t$) of
\begin{eqnarray}
F_{\rm s}(t, \Delta t) & = & \left| \frac{P_{\rm s}^{L}(t,t+\Delta
t) - P_{\rm s}^{R}(t,t+\Delta t)}{P_{\rm s}^{L}(t,t+\Delta t) +
P_{\rm s}^{R}(t,t+\Delta t)} \right| \nonumber
\\
& = & \left| \frac{\gamma_L e^{-\gamma_L t} - \gamma_R
e^{-\gamma_R t}}{\gamma_L e^{-\gamma_L t} + \gamma_R e^{-\gamma_R
t}} \right|.
\end{eqnarray}
If the detector switches soon after the measurement starts,
i.e.~much earlier than $\tau_0 \equiv
(\log\gamma_R-\log\gamma_L)/(\gamma_R-\gamma_L)$ [with $\log$
being the natural logarithm], the likely state of the qubit is
$\ket{R}$ with fidelity
\begin{equation}
F_{\rm s}(0, \Delta t) = \frac{\gamma_R - \gamma_L}{\gamma_R +
\gamma_L},
\label{Eq:Fidelity_switching}
\end{equation}
whereas if the detector switches at a time much later than
$\tau_0$, the likely state of the qubit is $\ket{L}$ with a
fidelity that approaches one. Interestingly, if the detector
switches at time $\tau_0$, the fidelity vanishes. In this case the
measurement does not give any information about the state of the
qubit, and the state of the qubit immediately after the switching
is equal to the initial state up to a simple rotation
\cite{ReversingTheMeasurement}. In other words, the purity of the
qubit's state does not change when this outcome is observed (note
that the qubit's state is, in general, a mixed state). The
function $F_{\rm s}(t, \Delta t)$ is plotted in Fig.~2 for the
case $\gamma_R/\gamma_L=10$.

\begin{figure}[h]
\includegraphics[width=7.0cm]{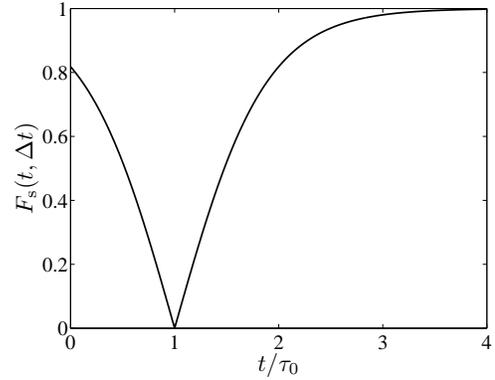}
\caption{The measurement fidelity $F_{\rm s}(t, \Delta t)$ as a
function of the switching time $t$ for $\gamma_R/\gamma_L=10$ and
$\beta=0$. The accuracy to which the switching time is known,
$\Delta t$, is assumed to be much smaller than $1/E$. Therefore,
the fidelity is independent of $\Delta t$ and is a function of $t$
only. Recall that $\tau_0 \equiv
(\log\gamma_R-\log\gamma_L)/(\gamma_R-\gamma_L)$.}
\end{figure}

\subsubsection{Overall fidelity}

The overall fidelity is now evaluated by taking the statistical
average over all possible outcomes
\begin{eqnarray}
F & = & \sum_j F_j {\rm Tr} \left\{ \hat{U}_j \rho_{\rm max.
mixed} \hat{U}_j^{\dagger} \right\} \nonumber
\\
& = & \int_0^{\infty} F_{\rm s}(t,dt) \frac{P_{\rm s}^{L}(t, t +
dt) + P_{\rm s}^{R}(t, t + dt)}{2} \nonumber
\\
& = & \frac{1}{2} \int_0^{\infty} \left| \frac{\gamma_R
e^{-\gamma_R t} - \gamma_L e^{-\gamma_L t}}{\gamma_R e^{-\gamma_R
t} + \gamma_L e^{-\gamma_L t}} \right| \times \nonumber
\\
& & \hspace{2cm} \left( \gamma_R e^{-\gamma_R t} + \gamma_L
e^{-\gamma_L t} \right) dt \nonumber
\\
& = & \frac{1}{2} \int_0^{\tau_0} \left(\gamma_R e^{-\gamma_R t} -
\gamma_L e^{-\gamma_L t} \right) dt + \nonumber
\\
& & \hspace{2cm} \frac{1}{2} \int_{\tau_0}^{\infty} \left(
\gamma_L e^{-\gamma_L t} - \gamma_R e^{-\gamma_R t} \right) dt
\nonumber
\\
& = & e^{- \gamma_L \tau_0} - e^{- \gamma_R \tau_0} \nonumber
\\
& = & \left( \frac{\gamma_R}{\gamma_L}
\right)^{-\frac{\gamma_L}{\gamma_R-\gamma_L}} - \left(
\frac{\gamma_R}{\gamma_L}
\right)^{-\frac{\gamma_R}{\gamma_R-\gamma_L}}.
\label{Eq:Overall_fidelity}
\end{eqnarray}
This expression for the fidelity is plotted as a function of
$\gamma_R/\gamma_L$ in Fig.~3. One could say that for large values
of $\gamma_R/\gamma_L$ one has a strong measurement, whereas small
values of $\gamma_R/\gamma_L$ correspond to weak measurement. As
can be seen from Fig.~2, however, even for small values of
$\gamma_R/\gamma_L$ there can be instances where a high-fidelity
outcome is obtained (when $t/\tau_0 \gg 1$).

\begin{figure}[h]
\includegraphics[width=7.0cm]{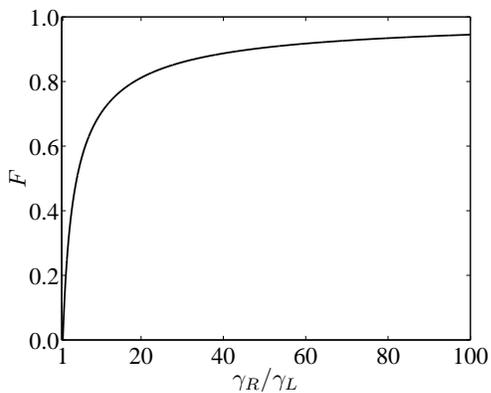}
\caption{Overall fidelity as a function of the ratio of switching
rates $\gamma_R/\gamma_L$ when the qubit Hamiltonian commutes with
the probed qubit operator, i.e.~in the special case $\beta=0$.}
\end{figure}

It is interesting to compare the above result for the overall
fidelity with the one that would be obtained in the following
situation: Let us assume that we connect the qubit to the detector
for some duration $\tau$ and only at the end check whether the
detector has switched or not. A switching instance would be
associated with the state $\ket{R}$ and a no-switching instance
would be associated with the state $\ket{L}$. The relevant
switching probabilities are given by:
\begin{eqnarray}
P_{\rm s}^{L}(0,\tau) & = & \left( 1 - e^{-\gamma_L \tau} \right)
\nonumber
\\
P_{\rm s}^{R}(0,\tau) & = & \left( 1 - e^{-\gamma_R \tau} \right).
\end{eqnarray}
The fidelities for switching and no-switching instances are
therefore given by
\begin{eqnarray}
F_{\rm s}(0, \tau) & = & \frac{e^{-\gamma_L \tau} - e^{-\gamma_R
\tau}}{2 - e^{-\gamma_L \tau} - e^{-\gamma_R \tau}} \nonumber
\\
F_{\rm ns}(\tau) & = & \frac{e^{-\gamma_L \tau} - e^{-\gamma_R
\tau}}{e^{-\gamma_L \tau} + e^{-\gamma_R \tau}},
\end{eqnarray}
and the overall fidelity is given by
\begin{eqnarray}
F & = & \frac{e^{-\gamma_L \tau} - e^{-\gamma_R \tau}}{2 -
e^{-\gamma_L \tau} - e^{-\gamma_R \tau}} \left( \frac{2 -
e^{-\gamma_L \tau} - e^{-\gamma_R \tau}}{2} \right) \nonumber
\\
& & \hspace{0cm} + \frac{e^{-\gamma_L \tau} - e^{-\gamma_R
\tau}}{e^{-\gamma_L \tau} + e^{-\gamma_R \tau}} \left(
\frac{e^{-\gamma_L \tau} +
e^{-\gamma_R \tau}}{2} \right) \nonumber \\
& = & e^{-\gamma_L \tau} - e^{-\gamma_R \tau}.
\end{eqnarray}
This expression for the overall fidelity is maximized by choosing
the pulse duration $\tau=\tau_0$ (which was defined in
Sec.~IV.A.1). This means that the optimal pulse duration is
determined by the switching time $\tau_0$ at which the fidelity
$F_{\rm s} (0,\tau_0)$ of Eq.~(\ref{Eq:Fidelity_switching})
vanishes, i.e.~the switching time that separates the two time
regions associated with the states $\ket{L}$ and $\ket{R}$. No
enhancement in overall fidelity is obtained by keeping track of
the exact switching time.

One thing that can be gained by keeping track of the exact
switching time is our knowledge about the state of the qubit after
the switching event in a specific experimental run. For example,
starting with a maximally mixed state and choosing instances
associated with high fidelity (e.g. those with no switching until
times much later than $\tau_0$), one can predict with a high
degree of certainty the qubit's state after the measurement.

\subsection{Case 2: $E \ll \gamma_-$ (Instantaneous measurement)}

This case is quite simple as it corresponds to the limit of
instantaneous measurement [note that the above condition
guarantees that $E \ll \gamma_R$; recall that
$\gamma_-=(\gamma_R-\gamma_L)/2$]. Equations
(\ref{Eq:lambda_and_phi}) now reduce to
\begin{eqnarray}
\lambda_{\pm} & \approx & - \frac{\gamma_+}{2} \pm \frac{1}{2}
\gamma_- \nonumber
\\
\eta & \approx & \beta,
\end{eqnarray}
such that
\begin{eqnarray}
\hat{U}_{\rm ns} (t) & = & \hat{U}_{\rm ns}^{\rm meas} (t)
\nonumber
\\
\hat{U}_{\rm s} (t,\Delta t) & = & \hat{U}_{\rm s}^{\rm meas}
(t,\Delta t) \nonumber
\\
\hat{U}_{\rm ns}^{\rm meas} (t) & = & e^{- \frac{\gamma_L}{2} t}
\ket{L} \bra{L} + e^{- \frac{\gamma_R}{2} t} \ket{R} \bra{R}
\nonumber
\\
\hat{U}_{\rm s}^{\rm meas} (t,\Delta t) & = & \sqrt{\gamma_L
\Delta t} \; e^{- \frac{\gamma_L}{2} t} \ket{L} \bra{L} \nonumber \\
& & \hspace{0.5cm} + \sqrt{\gamma_R \Delta t} \; e^{-
\frac{\gamma_R}{2} t} \ket{R} \bra{R}.
\label{Eq:Us_instantaneous_measurement}
\end{eqnarray}
Note that the energy $E$ does not appear in the above expressions;
the detector switches so fast, at least for one of the qubit
states, that the measurement is completed before the qubit
Hamiltonian contributes any change to the state of the qubit.

The fact that $E$ can be ignored in this case is similar to the
situation analyzed in Sec.~IV.A (case 1). One can also see that
the last two lines in Eq.~(\ref{Eq:Us_instantaneous_measurement})
coincide with Eq.~(\ref{Eq:Umeas_zero_beta}). One can therefore
follow the analysis of Sec.~IV.A [starting from
Eq.~(\ref{Eq:Umeas_zero_beta})] in order to calculate the
measurement fidelity for the instantaneous-measurement case under
the different possible scenarios. Not surprisingly, the
measurement basis is $\{ \ket{L} , \ket{R} \}$, which is natural
given the fact that the measurement is instantaneous, and the
measurement basis should coincide with the natural measurement
basis of the detector.

As in Sec.~IV.A, the overall fidelity is given by
Eq.~(\ref{Eq:Overall_fidelity}) [which is plotted in Fig.~3]. In
the limit $\gamma_R/\gamma_L \rightarrow \infty$, the fidelity is
one, and one obtains the simple case of a strong (projective)
measurement in the basis $\left\{ \ket{L}, \ket{R} \right\}$. Note
that the limit $\gamma_R/\gamma_L \rightarrow 0$ gives the same
result, with the roles of $\gamma_L$ and $\gamma_R$ reversed.

In the limit $\gamma_- \ll \gamma_+$ (i.e.~when
$\gamma_R/\gamma_L$ is close to one), we still have an
instantaneous measurement, but with a low fidelity. The
measurement in this limit is therefore a weak measurement. The
switching time of the detector allows the experimenter to make a
guess about the state of the qubit, but with a small degree of
confidence. Repeated measurements of this type, possibly
synchronized with the qubit's free precession, on a single qubit
would increase the acquired amount of information (or degree of
certainty) about the state of the qubit, thus approaching the case
of a strong measurement.

\subsection{Case 3: $E \gg \gamma_+$ (Slow measurement)}

The opposite limit of case 2 is the one with $E \gg \gamma_-$.
However, when $\gamma_- \ll E \ll \gamma_+$, we recover the case
of an instantaneous, weak measurement (with minor changes).
Therefore, in this subsection we focus on the case where $E$ is
larger than both $\gamma_-$ and $\gamma_+$.

When $E \gg \gamma_-$, we find that (see Appendix A)
\begin{widetext}
\begin{eqnarray}
\hat{U}_{\rm ns}^{\dagger} (t) \hat{U}_{\rm ns} (t) & = & \left(
\begin{array}{cc}
e^{-(\gamma_+ - \gamma_-\cos\beta) t} & 0 \\
0 & e^{-(\gamma_+ + \gamma_-\cos\beta) t}
\end{array}
\right) \nonumber \\
\hat{U}_{\rm s}^{\dagger} (t,\Delta t) \hat{U}_{\rm s} (t,\Delta
t) & = & P_{\rm s}^{(1)}(t,\Delta t) \ket{\psi_{\rm s}^{(1)}(t)}
\bra{\psi_{\rm s}^{(1)}(t)} + P_{\rm s}^{(2)}(t,\Delta t)
\ket{\psi_{\rm s}^{(2)}(t)} \bra{\psi_{\rm s}^{(2)}(t)}
\label{Eq:UdaggerU_for_case_3}
\end{eqnarray}
where
\begin{eqnarray}
\ket{\psi_{\rm s}^{(1)}(t)} & = & \left(
\begin{array}{c}
\cos \frac{\theta(t)}{2} e^{- i \frac{\tilde{E}}{2} t} \\
\sin \frac{\theta(t)}{2} e^{i \frac{\tilde{E}}{2} t} \\
\end{array}
\right) = \left(
\begin{array}{cc}
e^{- i \frac{\tilde{E}}{2} t} & 0 \\
0 & e^{i \frac{\tilde{E}}{2} t} \\
\end{array}
\right) \left(
\begin{array}{c}
\cos \frac{\theta(t)}{2}\\
\sin \frac{\theta(t)}{2}\\
\end{array}
\right) \nonumber
\\
\ket{\psi_{\rm s}^{(2)}(t)} & = & \left(
\begin{array}{c}
\sin \frac{\theta(t)}{2} e^{- i \frac{\tilde{E}}{2} t} \\
-\cos \frac{\theta(t)}{2} e^{i \frac{\tilde{E}}{2} t} \\
\end{array}
\right) = \left(
\begin{array}{cc}
e^{- i \frac{\tilde{E}}{2} t} & 0 \\
0 & e^{i \frac{\tilde{E}}{2} t} \\
\end{array}
\right) \left(
\begin{array}{c}
\sin \frac{\theta(t)}{2}\\
-\cos \frac{\theta(t)}{2}\\
\end{array}
\right) \nonumber
\\
\tilde{E} & = & E-\gamma_-^2/2E \nonumber
\end{eqnarray}
\begin{eqnarray}
\tan \theta(t) & = & \frac{(\gamma_L-\gamma_R) \sin\beta}{\gamma_L
\left( e^{\gamma_-t \cos\beta} \cos^2\frac{\beta}{2} -
e^{-\gamma_-t \cos\beta} \sin^2\frac{\beta}{2} \right) + \gamma_R
\left( e^{\gamma_-t \cos\beta} \sin^2\frac{\beta}{2} -
e^{-\gamma_-t \cos\beta} \cos^2\frac{\beta}{2} \right)} \nonumber
\\
P_{\rm s}^{(1)}(t,\Delta t) + P_{\rm s}^{(2)}(t,\Delta t) & = &
e^{-\gamma_+ t} \Delta t \nonumber \\ & & \left\{ \gamma_L \left(
e^{\gamma_-t \cos\beta} \cos^2\frac{\beta}{2} + e^{-\gamma_-t
\cos\beta} \sin^2\frac{\beta}{2} \right) + \gamma_R \left(
e^{\gamma_-t \cos\beta} \sin^2\frac{\beta}{2} + e^{-\gamma_-t
\cos\beta} \cos^2\frac{\beta}{2} \right) \right\} \nonumber
\\
P_{\rm s}^{(1)}(t,\Delta t) - P_{\rm s}^{(2)}(t,\Delta t) & = &
e^{-\gamma_+ t} \Delta t \frac{(\gamma_L-\gamma_R)
\sin\beta}{\sin\theta(t)}.
\label{Eq:Measurement_Basis}
\end{eqnarray}
\end{widetext}
Recall that, in spite of their appearance, the operators
$\hat{U}_{\rm ns}(t)$ and $\hat{U}_{\rm s}(t,\Delta t)$ are not
unitary matrices.

For short times ($t=0$), the above expressions reduce to
\begin{eqnarray}
\theta & = & \beta \nonumber
\\
P_{\rm s}^{(1)} + P_{\rm s}^{(2)} & = & \Delta t \left( \gamma_L +
\gamma_R \right) \nonumber
\\
P_{\rm s}^{(1)} - P_{\rm s}^{(2)} & = & \Delta t
(\gamma_L-\gamma_R) \nonumber
\\
F_{\rm s}(0,\Delta t) & = &
\frac{\gamma_R-\gamma_L}{\gamma_R+\gamma_L},
\end{eqnarray}
such that the measurement basis coincides with the detector's
natural measurement basis, as it should for short times (where the
qubit Hamiltonian has not affected the qubit's state). For long
times ($\gamma_-t\rightarrow\infty$) [and $\beta \neq \pi/2$], on
the other hand, the above expressions reduce to
\begin{widetext}
\begin{eqnarray}
\theta & \rightarrow & 0 \nonumber
\\
P_{\rm s}^{(1)} + P_{\rm s}^{(2)} & = & e^{-\gamma_+ t} \Delta t
\left\{ \gamma_L e^{\gamma_-t \cos\beta} \cos^2\frac{\beta}{2} +
\gamma_R e^{\gamma_-t \cos\beta} \sin^2\frac{\beta}{2} \right\}
\nonumber
\\
P_{\rm s}^{(1)} - P_{\rm s}^{(2)} & = & e^{-\gamma_+ t} \Delta t
\left\{ \gamma_L e^{\gamma_-t \cos\beta} \cos^2\frac{\beta}{2} +
\gamma_R e^{\gamma_-t \cos\beta} \sin^2\frac{\beta}{2} \right\}
\nonumber
\\
F(t\rightarrow\infty,\Delta t) & = & 1,
\end{eqnarray}
\end{widetext}
such that the measurement is performed in the qubit's energy
eigenbasis with 100\% fidelity. Thus the measurement basis starts
from the detector's natural measurement basis (i.e.~$\left\{
\ket{L}, \ket{R} \right\}$) at $t=0$ and spirals towards the $z$
axis (i.e.~$\left\{ \ket{0}, \ket{1} \right\}$) with time. Note
that since the denominator in the first line of
Eq.~(\ref{Eq:Measurement_Basis}) goes from negative values
(e.g.~at $t=0$) to positive values (e.g.~when
$t\rightarrow\infty$) the measurement basis goes through the
equator during its spiraling motion.

From the above results, we can see that the measurement is made in
a rotated basis that is determined by the exact switching time,
which is uncontrollable. Each experimental run gives a measurement
outcome that is different from other identically prepared runs.
Note, in particular, that when $\gamma_L=0$ the fidelity is 100\%
for all switching times $t$ (which can be verified with
straightforward algebra). This result is quite interesting; even
though the fidelity is 100\% for every single run, the measurement
basis is unpredictable and is only determined when the switching
event occurs.

Let us now consider the situation where we know only whether the
detector switched between $t=0$ and $t=\tau$ (with
$\tilde{E}\tau\gg 1$) or not. The measurement matrices that
correspond to these two possible outcomes are given by
\begin{eqnarray}
\hat{U}_{\rm s}^{\dagger}(0,\tau) \hat{U}_{\rm s}(0,\tau) & = &
\int_0^{\tau} \hat{U}_{\rm s}^{\dagger}(t,dt) \hat{U}_{\rm
s}(t,dt) \nonumber \\
& = & 1 - \hat{U}_{\rm ns}^{\dagger}(\tau) \hat{U}_{\rm ns}(\tau),
\end{eqnarray}
and $\hat{U}_{\rm ns}^{\dagger}(\tau) \hat{U}_{\rm ns}(\tau)$ as
given in Eq.~(\ref{Eq:UdaggerU_for_case_3}). Since the measurement
matrices are diagonal in the qubit's energy eigenbasis, the
measurement is performed in that basis. The overall fidelity can
be calculated similarly to what was done in Sec.~IV.A:
\begin{equation}
F = e^{-(\gamma_+-\gamma_-\cos\beta) \tau} -
e^{-(\gamma_++\gamma_-\cos\beta) \tau}.
\end{equation}
This expression is maximized by choosing
\begin{equation}
\tau = \frac{\log
\frac{\gamma_++\gamma_-\cos\beta}{\gamma_+-\gamma_-\cos\beta}}
{2\gamma_- \cos\beta},
\end{equation}
which gives the maximum fidelity
\begin{eqnarray}
F_{\rm max} & = & \left(
\frac{\gamma_+-\gamma_-\cos\beta}{\gamma_++\gamma_-\cos\beta}
\right)^{\frac{\gamma_+-\gamma_-\cos\beta}{2\gamma_-\cos\beta}}
\nonumber \\ & & \hspace{1cm} - \left(
\frac{\gamma_+-\gamma_-\cos\beta}{\gamma_++\gamma_-\cos\beta}
\right)^{\frac{\gamma_++\gamma_-\cos\beta}{2\gamma_-\cos\beta}}.
\end{eqnarray}
Since the above expression looks somewhat cumbersome, we shall not
pursue it in its general form any further. In the special case
$\gamma_L=0$, the overall fidelity for a readout pulse of duration
$\tau$ is given by
\begin{equation}
F = e^{-\gamma_R \tau \sin^2\frac{\beta}{2}}  - e^{-\gamma_R \tau
\cos^2\frac{\beta}{2}},
\end{equation}
which is maximized by choosing
\begin{equation}
\tau = \frac{- \log \tan^2\frac{\beta}{2}}{\gamma_R \cos\beta}.
\end{equation}
The maximum fidelity is therefore given by
\begin{eqnarray}
F_{\rm max} & = &
\left(\tan\frac{\beta}{2}\right)^{\frac{2\sin^2(\beta/2)}{\cos\beta}}
-
\left(\tan\frac{\beta}{2}\right)^{\frac{2\cos^2(\beta/2)}{\cos\beta}}
\nonumber
\\
& = & \left(\tan\frac{\beta}{2}\right)^{\sec\beta-1} -
\left(\tan\frac{\beta}{2}\right)^{\sec\beta+1}.
\label{Eq:Fidelity_incoherent}
\end{eqnarray}
This expression is plotted in Fig.~4. The measurement fidelity
decreases as $\beta$ approaches $\pi/2$ and vanishes at that
point.

\begin{figure}[h]
\includegraphics[width=7.0cm]{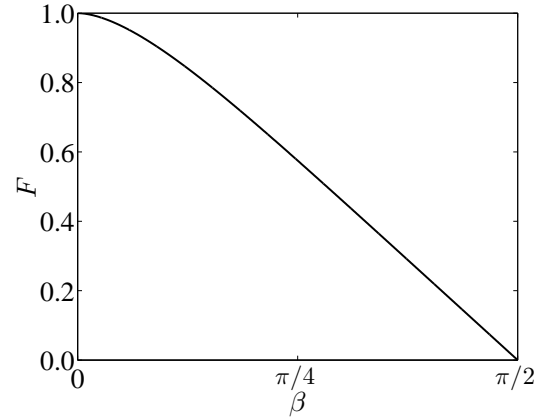}
\caption{Maximum fidelity as a function of the angle between the
qubit Hamiltonian and measurement axes, assuming that the exact
switching time is not known (up to the measurement duration, which
is taken to be much larger than $1/E$).}
\end{figure}

For further demonstration of the results discussed in this
section, see Appendix B, where the special case $\beta=\pi/2$ is
analyzed in some detail.

\subsection{All-in-one measurement setting}

Looking at the results of Sec.~IV.C (case 3), it is interesting
that every time the experiment is repeated, the measurement
outcome gives information about the qubit state in a different
basis (that cannot be predicted beforehand). This result suggests
that it could be possible to perform `single-setting quantum state
tomography' \cite{Single_setting}; instead of the externally
applied rotations used to perform measurements in different bases
in conventional quantum state tomography, the (uncontrollable)
switching times determine the measurement basis in each
experimental run. Indeed, using
Eq.~(\ref{Eq:UdaggerU_for_switching}) we find that the switching
probability as a function of time is given by
%
\begin{eqnarray}
{\rm Prob_s}(t,\Delta t) & = & {\rm Tr} \left\{ \hat{U}_{\rm
s}^{\dagger} (t,\Delta t) \hat{U}_{\rm s} (t,\Delta t)
\rho(0) \right\} \nonumber \\
& = & e^{-\gamma_+ t} \Delta t \Bigg[ \rho_{00} e^{\gamma_-t
\cos\beta} \left( \gamma_+ - \gamma_- \cos\beta \right) \nonumber
\\ & & \hspace{0cm} + \rho_{11} e^{-\gamma_-t \cos\beta} \left(
\gamma_+ + \gamma_- \cos\beta \right) \nonumber \\ & &
\hspace{0cm} - \left( \rho_{10}+\rho_{01}
\right) \frac{\gamma_-}{4} \sin\beta \cos \tilde{E}t \nonumber \\
& & \hspace{0cm} - \left( \frac{ \rho_{10}-\rho_{01}}{i} \right)
\frac{\gamma_-}{4} \sin\beta \sin \tilde{E}t \Bigg].
\end{eqnarray}
%
By fitting the experimentally measurable switching-time
probability distribution to the above function, one can extract
the three quantities $\left( \rho_{11}-\rho_{00} \right)$, $\left(
\rho_{10}+\rho_{01} \right)$ and $\left( \rho_{10}-\rho_{01}
\right)/i$, which completely characterize the quantum state of the
qubit. In addition to the above three parameters, one can also
extract the values of $\gamma_L$, $\gamma_R$, $\beta$ and $E$,
thus providing an all-in-one measurement that characterizes both
the system parameters as well as the quantum state of the qubit
\cite{Aharonov}. Note that in the special cases $\beta=0$,
$\beta=\pi/2$ and $\gamma_+ \gg E$ full system characterization
and state tomography are not possible, in agreement with intuitive
expectations. In particular, the case $\beta=0$ corresponds to a
measurement in the $\{\ket{0}, \ket{1} \}$ basis (regardless of
the switching time), which clearly cannot give any information
about the quantities $\left( \rho_{10}+\rho_{01} \right)$ and
$\left( \rho_{10}-\rho_{01} \right)/i$.

\section{Coherent detector}

In the preceding two sections, we have focused on the regime where
the coherence time between the two detector states (i.e.~the
switched and non-switched states $\ket{S}$ and $\ket{N}$) is the
shortest timescale in the problem. Now we turn to the case where
this time is longer than $1/E$, which now is the shortest
timescale in the problem.

We should emphasize here the distinction between this case of a
coherent detector and the case of a `slow' but incoherent detector
(with a small switching rate; $\gamma_+ \ll E$) analyzed in the
previous section \cite{Martin}. That situation was analyzed using
the same theoretical framework as any other case with an
incoherent detector. We should also emphasize that, in spite of
the detector's long coherence time (compared to $1/E$), we still
assume irreversible switching. These assumptions are compatible;
for irreversibility one only needs to assume that the detector's
coherence time is much shorter than the inverse of the switching
rate. The assumptions explained above can be summarized as
follows:
\begin{equation}
\frac{1}{E} \ll {\rm coh. \ time \ of \ detector} \ll \frac{1}{\rm
switching \ rate}.
\end{equation}

Since we are now treating the detector as a quantum-coherent
object (at least on the timescale of qubit precession), we analyze
the dynamics using a Hamiltonian that describes the combined
qubit-detector system. Since the Hamiltonian contains the physical
interaction between the qubit and the detector, in this section we
find it easier to define the qubit's $z$ axis as defined by the
probed operator. Thus, the Hamiltonian can be expressed as
\begin{equation}
\hat{H} = \hat{H}_{\rm q} + \hat{H}_{\rm det},
\end{equation}
where
\begin{eqnarray}
\hat{H}_{\rm q} & = & -\frac{E}{2} \left( \sin\beta
\hat{\sigma}_x + \cos\beta \hat{\sigma}_z \right) \nonumber \\
\hat{H}_{\rm det} & = & \left\{ g_L \left( \ket{S}\bra{N} + h.c.
\right) + \epsilon_L \ket{S}\bra{S} \right\} \ket{L} \bra{L} + \nonumber \\
& & \left\{ g_R \left( \ket{S}\bra{N} + h.c. \right) + \epsilon_R
\ket{S}\bra{S} \right\} \ket{R} \bra{R},
\end{eqnarray}
$g_L$ and $g_R$ are the coupling strengths that induce detector
switching for the qubit states $|L\rangle$ and $|R\rangle$,
respectively, and $\epsilon_L$ and $\epsilon_R$ are relative
energy biases between the states $\ket{N}$ and $\ket{S}$ for the
two qubit states. As before, the detector is assumed to be
initially in the state $\ket{N}$.

Since the qubit's energy is the largest energy scale in the
problem, the dynamics described by the above Hamiltonian and the
detector's decoherence cannot induce transitions between the
qubit's ground and excited states (note also that we are ignoring
any intrinsic decoherence mechanisms of the qubit). We can
therefore analyze the dynamics by treating two separate subspaces
of the Hilbert space, one with the qubit in the ground state
$\ket{0}$ and one with the qubit in the excited state $\ket{1}$.
This approach results in two effective Hamiltonians for the
detector:
\begin{eqnarray}
\hat{H}_{{\rm det},\ket{0}} & = & {}_{\rm qubit} \hspace{-0.05cm}
\bra{0} \hat{H} \ket{0}_{\rm qubit}
\nonumber \\
& = & \bar{g}_0 \left( \ket{S}\bra{N} + h.c. \right) +
\bar{\epsilon}_0 \ket{S}\bra{S} \nonumber
\\
\hat{H}_{{\rm det},\ket{1}} & = & {}_{\rm qubit} \hspace{-0.05cm}
\bra{1} \hat{H} \ket{1}_{\rm qubit}
\nonumber \\
& = & \bar{g}_1 \left( \ket{S}\bra{N} + h.c. \right) +
\bar{\epsilon}_1 \ket{S}\bra{S},
\label{Eq:Detector_subspace_Hamiltonian}
\end{eqnarray}
where
\begin{eqnarray}
\bar{g}_0 & = & g_L \cos^2\frac{\beta}{2} + g_R
\sin^2\frac{\beta}{2} \nonumber
\\
\bar{g}_1 & = & g_L \sin^2\frac{\beta}{2} + g_R
\cos^2\frac{\beta}{2} \nonumber
\\
\bar{\epsilon}_0 & = & \epsilon_L \cos^2\frac{\beta}{2} +
\epsilon_R \sin^2\frac{\beta}{2} \nonumber
\\
\bar{\epsilon}_1 & = & \epsilon_L \sin^2\frac{\beta}{2} +
\epsilon_R \cos^2\frac{\beta}{2}.
\end{eqnarray}

In order to avoid running into complicated expressions below, we
make the experimentally relevant assumption that the switching for
the state $\ket{R}$ is much faster than that for the state
$\ket{L}$. This situation occurs when one or both of the following
conditions are satisfied: (1) $|g_R| \gg |g_L|$ or (2)
$|\epsilon_L|$ is much larger than all other parameters relevant
to the switching dynamics. We analyze these two cases separately.

First we consider the case where $|g_R| \gg |g_L|$, such that
$g_L$ can be ignored in the following analysis. For simplicity and
clarity, we take $\epsilon_L$ and $\epsilon_R$ to be smaller than
the detector's coherence time, such that they also can be ignored
for purposes of this calculation. Equations
(\ref{Eq:Detector_subspace_Hamiltonian}) therefore reduce to
\begin{eqnarray}
\hat{H}_{{\rm det},\ket{0}} & = & g_R \sin^2\frac{\beta}{2} \left(
\ket{S}\bra{N} + h.c. \right) \nonumber
\\
\hat{H}_{{\rm det},\ket{1}} & = & g_R \cos^2\frac{\beta}{2} \left(
\ket{S}\bra{N} + h.c. \right).
\end{eqnarray}
Since we are assuming that the switching is an incoherent process
(or, in other words, the detector's coherence time is much shorter
than $|g_R|$), the switching rates are proportional to the squares
of the transition coefficients in the above Hamiltonians. The
switching rates for the qubit states $\ket{0}$ and $\ket{1}$ can
therefore be expressed as
\begin{eqnarray}
\Gamma_0 & = & \Gamma_R \sin^4\frac{\beta}{2} \nonumber \\
\Gamma_1 & = & \Gamma_R \cos^4\frac{\beta}{2}.
\label{Eq:Rates_coherent_detector_1}
\end{eqnarray}

We now consider the second case, where $|\epsilon_L|$ is very
large compared to all other parameters relevant to the switching
dynamics. In this case the switching rates are governed by the
scaling formula
\begin{equation}
\Gamma_j \propto \frac{\bar{g}_j^2}{|\bar{\epsilon}_j|}.
\end{equation}
With the simplifying assumption that $g_L=g_R$, we find switching
rates of the form
\begin{eqnarray}
\Gamma_0 & = & \Gamma_R \sec^2\frac{\beta}{2} \nonumber \\
\Gamma_1 & = & \Gamma_R \csc^2\frac{\beta}{2}.
\label{Eq:Rates_coherent_detector_2}
\end{eqnarray}
Note that the divergence of $\Gamma_1$ for $\beta\rightarrow 0$ is
an artefact of our approximations. In a more accurate description,
$\Gamma_1$ would saturate at a finite value as $\beta$ approaches
zero.

The difference between the expressions in
Eqs.~(\ref{Eq:Rates_coherent_detector_1}) and
(\ref{Eq:Rates_coherent_detector_2}) shows that, in the case of a
coherent detector, knowledge of the microscopic details of the
detector is necessary in order to determine the dependence of the
switching rates on the angle $\beta$. On the other hand, this
sensitivity to details could be useful for experimentally testing
theoretical models of the detector.

One common, and important, result for the different cases with a
coherent detector, however, can be seen by noting that the qubit's
state in the energy eigenbasis $\left\{ \ket{0}, \ket{1} \right\}$
does not change during the measurement. Therefore, the detector
cannot be performing a measurement on the qubit in any basis other
than this one. Since, in addition, the detector's switching rate
depends on the qubit's state in the energy eigenbasis, one can
conclude that the detector does in fact perform a measurement on
the qubit in the energy eigenbasis. Using the transition rates
given in Eq.~(\ref{Eq:Rates_coherent_detector_1}) or
(\ref{Eq:Rates_coherent_detector_2}), one can follow the analysis
of Sec.~IV.A and determine the fidelity of the measurement setup.
The two expressions found in this section describe a fidelity that
slowly decreases from unity at $\beta=0$ to zero at $\beta=\pi/2$.
This result seems to be unavoidable for the two-state-detector
model considered so far. In spite of the different functional
dependence, the fidelity decrease is somewhat similar to the one
obtained in Sec.~IV.C for an incoherent detector with small
switching rates [see Eq.~(\ref{Eq:Fidelity_incoherent})]. In
Sec.~VII, we shall discuss a system where the fidelity of a
coherent detector remains very close to unity even as $\beta$
approaches $\pi/2$.

\section{Comparison between different types of detectors}

To conclude the above analysis of the detector's operation, we
present in Table I the measurement basis and the factors limiting
the measurement fidelity in different operation regimes.

\begin{center}
\begin{table}
\begin{tabular}{|l|l|l|} \hline
Detector type               & Measurement                        & Meas. fidelity \\
                            & basis                              & limited by \\
\hline \hline
Fast switching              & $\left\{ \ket{L},\ket{R} \right\}$ & $\gamma_R/\gamma_L$ \\
\hline
Slow switching,             & determined                         & $\gamma_R/\gamma_L$ \\
incoherent, switching time  & upon switching                     & \\
accurately measurable       &                                    & \\
\hline
Slow switching,             & $\left\{ \ket{0},\ket{1} \right\}$ & $\gamma_R/\gamma_L$ and $\beta$ \\
incoherent, exact switching &                                    & \\
time inaccessible           &                                    & \\
\hline
Slow switching,             & $\left\{ \ket{0},\ket{1} \right\}$ & $\gamma_R/\gamma_L$ and $\beta$ \\
coherent                    &                                    & (in two-state \\
                            &                                    & model; Sec.~V)    \\
\hline
\end{tabular}
\newline
Table I: Measurement basis and fidelity for different types of
detectors characterized by their coupling strength to the qubit,
their coherence times and whether the switching time can be known
accurately or not. In the case of a weakly coupled coherent
detector, we give the results for the two-state-detector model
only.
\end{table}
\end{center}

It is worth mentioning here a distinction between two different
types of weak measurement, i.e.~a measurement with low fidelity.
There are cases where the interaction between the qubit and the
detector is such that it is fundamentally impossible for any
observer (with full access to all degrees of freedom of the
detector) to determine with certainty the state of the qubit. The
situations analyzed in Secs.~IV.A, IV.B and V correspond to this
type of weak measurement. In this case, there is no lost
information, and an initially pure state of the qubit remains pure
after the measurement (note that not knowing the exact switching
time to an accuracy $\Delta t$ that is much smaller than $1/E$
does not constitute lost, and relevant, information in this
context; Going to higher accuracies within the regime $E \Delta t
\ll 1$ provides more information about the state of the detector,
but not that of the qubit). The situation where the observer
cannot establish the exact switching time of an incoherent
detector to an accuracy smaller than $1/E$ (analyzed in Sec.~IV.C)
involves information that is collected by the detector and
potentially measurable, but inaccessible to the observer in the
experimental situation under consideration. In this case, the
qubit's state after the measurement is generally a mixed state
(obtained by averaging over all possible outcomes within the
experimenter's accuracy range), even if the initial state is pure.
One could say that the accessible information serves as
measurement data, whereas information that is collected by the
detector but is inaccessible to the experimenter plays the role of
a decoherence mechanism.

We now pause for a moment to comment on the issue of qubit
decoherence. In some experimental situations, the detector
enhances the qubit's decoherence rate \cite{Makhlin}. This
additional decoherence channel typically causes loss of coherence
in the $\{ \ket{L}, \ket{R} \}$ basis. If the decoherence rate is
larger than the qubit's energy scale $E$, the qubit's state will
be frozen in the $\{ \ket{L}, \ket{R} \}$ basis during the
measurement. In that case, the detector's operation resembles that
in the instantaneous (and possibly strong-measurement) regime,
even if the switching rates are small compared to $E$. Thus, such
an additional decoherence could be detected by analyzing the
operation of the detector and comparing it with the expected
behaviour in the absence of decoherence \cite{Picot}. It is also
worth mentioning here that including decoherence into our
theoretical analysis would be quite nontrivial. The evolution of
the qubit's state during the short interval $\delta t$ would have
to be described by the sum of terms, and not just a single
propagator as was the case in our analysis. It would therefore not
be possible to derive a simple analytic expression for the matrix
$\hat{U}_{\rm ns}(t)$ in the presence of decoherence as we were
able to do in Eq.~(\ref{Eq:U_no_switch}).

To conclude the discussion of the two-state model, we summarize
some of the main principles that have been demonstrated in the
above analysis. For a strong measurement, the detector always
measured the qubit in the $\{ \ket{L}, \ket{R} \}$ basis. In the
case of a slow, incoherent detector whose switching time can be
determined accurately, one could say that, to a lowest
approximation, the qubit's state precesses freely following the
qubit Hamiltonian and is then measured in the $\{ \ket{L}, \ket{R}
\}$ basis when the detector switches. When the switching time can
only be known with small accuracy (i.e.~much longer than the
qubit's precession period), the detector effectively probes the
qubit in the energy eigenbasis, with the corresponding switching
rates being obtained as weighted averages of the bare switching
rates $\gamma_L$ and $\gamma_R$. A slow, coherent detector
measures the qubit in the energy eigenbasis, with the detector
effectively following weighted-average Hamiltonians for the two
energy eigenstates.

\section{Application to superconducting flux qubits}

In the preceding sections we have analyzed a theoretical model
where the detector is a two-state system. We now turn to a
specific physical example of switching-based measurement, namely
the readout of a superconducting flux qubit
\cite{OtherQubitTypes}. Although the two-state-detector model does
not provide an accurate description of the complex circuits that
we discuss in this section, the insight developed in the preceding
sections (in particular the general principles summarized in
Sec.~VI) will allow us to infer a number of results concerning the
realistic situations that we consider here.

There are two qubit-readout methods used in experiment that rely
on a switching process in a superconducting circuit: switching
from a zero-voltage to a finite-voltage state across a dc
superconducting quantum interference device (SQUID)
\cite{VanderWal} and switching between two possible dynamical
states of a nonlinear oscillator \cite{Siddiqi} (For a theoretical
study on measurement in superconducting qubits, see
e.g.~\cite{Johansson}; and for studies on quantum tunneling
between dynamical states in superconducting systems, see
e.g.~\cite{Marthaler,Serban,Nation}). Both of the readout methods
mentioned above can be modelled theoretically as a fictitious
particle moving in a trapping potential. Therefore, their
behaviour in terms of switching dynamics should exhibit the same
features, which we discuss below.

We briefly explain the operation of the dc-SQUID detector that
relies on switching from the zero-voltage state to the
finite-voltage state of a current biased dc-SQUID (see also
\cite{Nakano,MaassenVDB}); as mentioned above, the analysis of the
dynamical bifurcation would be similar. The principle of the
measurement process is as follows: The dc-SQUID is known to be a
highly sensitive device for measuring magnetic fields. In the
presence of an externally applied magnetic flux and a bias current
that is close to a certain (critical) value, the switching rate of
the dc-SQUID from the zero-voltage to the finite-voltage state
varies greatly with even small changes in magnetic field. When a
flux qubit is placed next to a dc-SQUID biased close to the
switching point, the clockwise- and counterclockwise-current
states of the qubit can result in two switching rates that are
different by orders of magnitude. By choosing suitable values for
the applied bias current and the length of the current pulse, the
SQUID switches with high probability for one qubit state and with
low probability for the other qubit state. The contrast between
the two switching probabilities is perhaps most clearly visualized
in the so-called S curves. These curves are produced by plotting
the SQUID's switching probability as a function of bias current
for the qubit's clockwise- and counterclockwise-current states
(for a fixed measurement duration). An example of such S curves is
shown in Fig.~5 (dashed lines).

\begin{figure}[h]
\includegraphics[width=6.0cm]{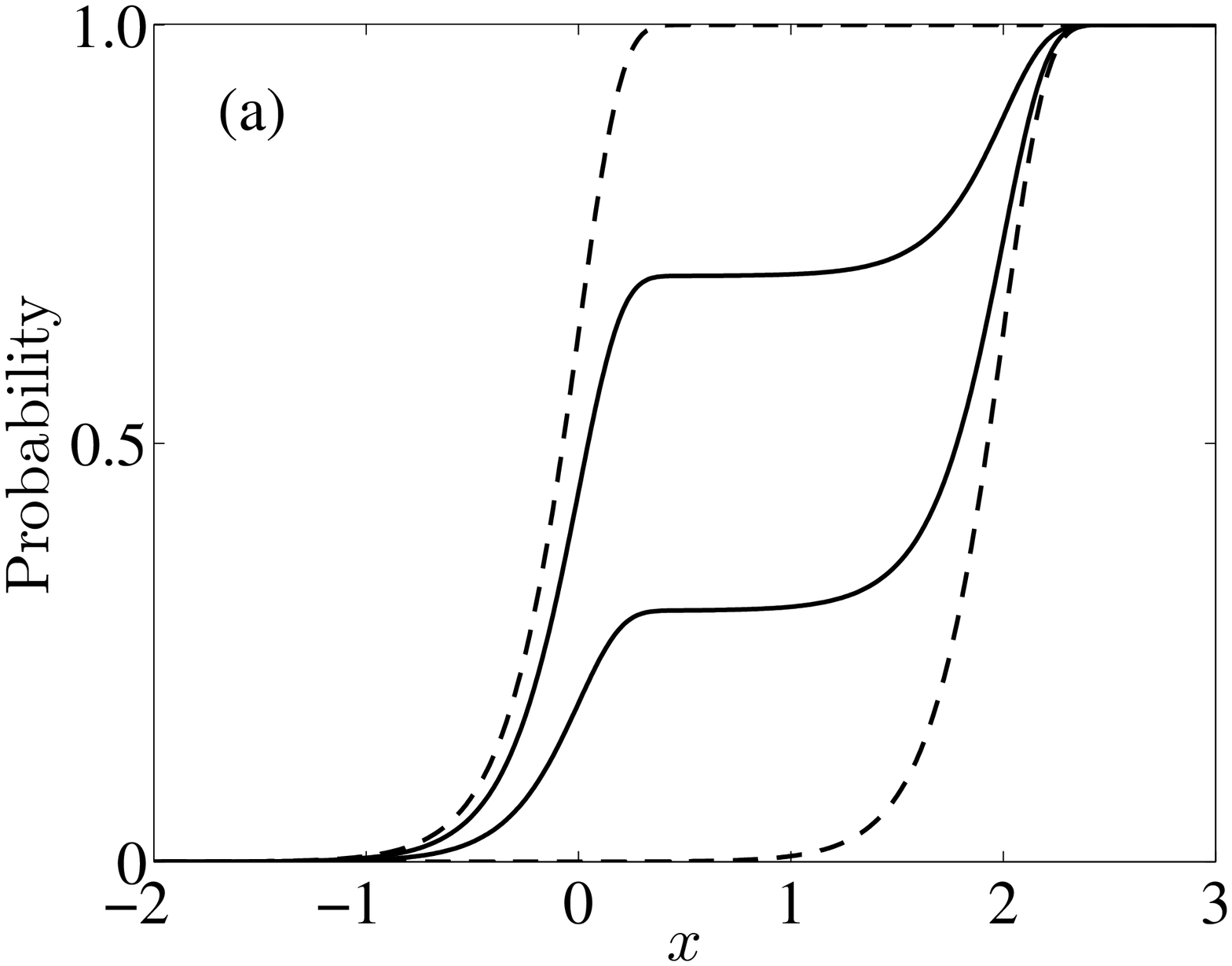}
\includegraphics[width=6.0cm]{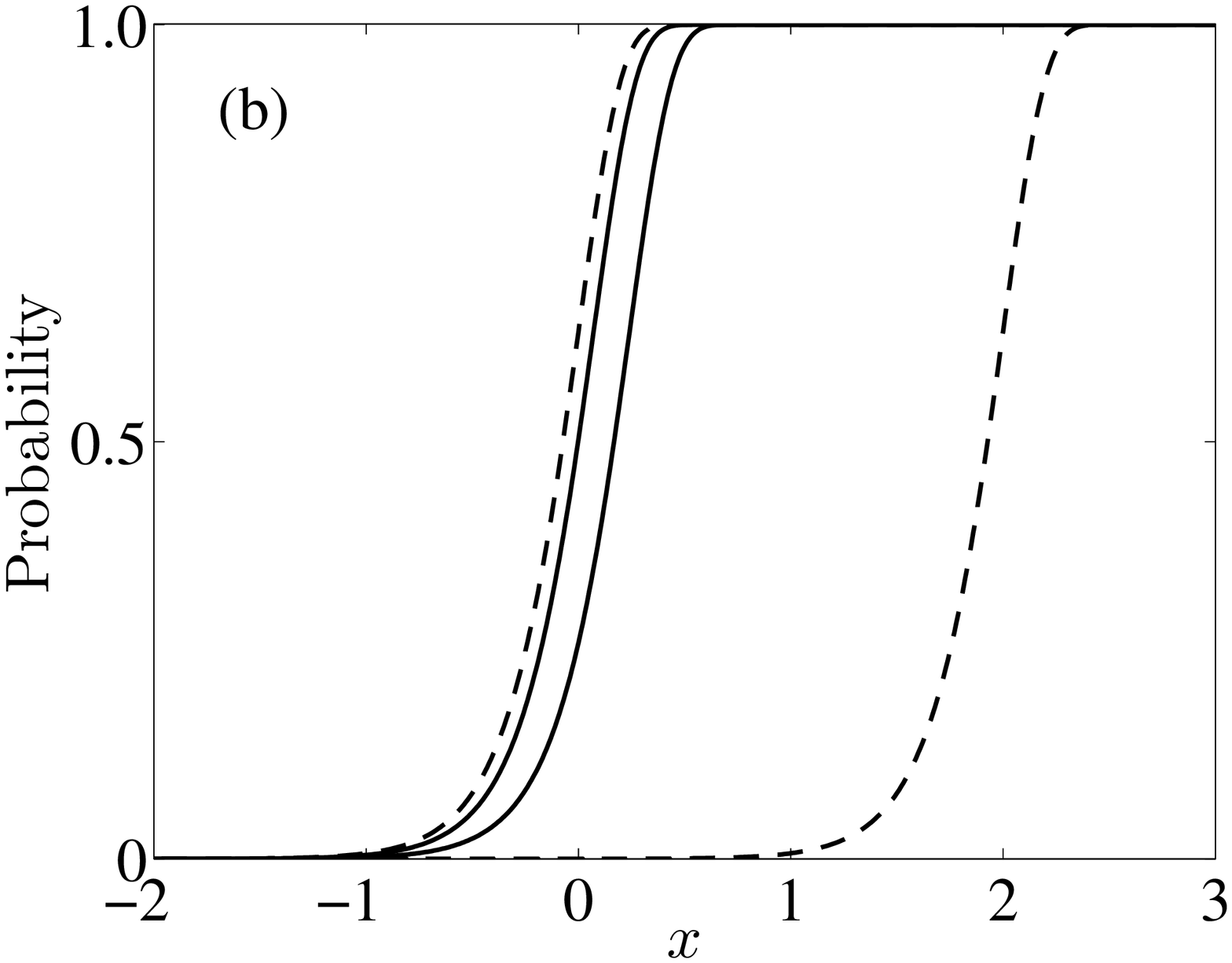}
\includegraphics[width=6.0cm]{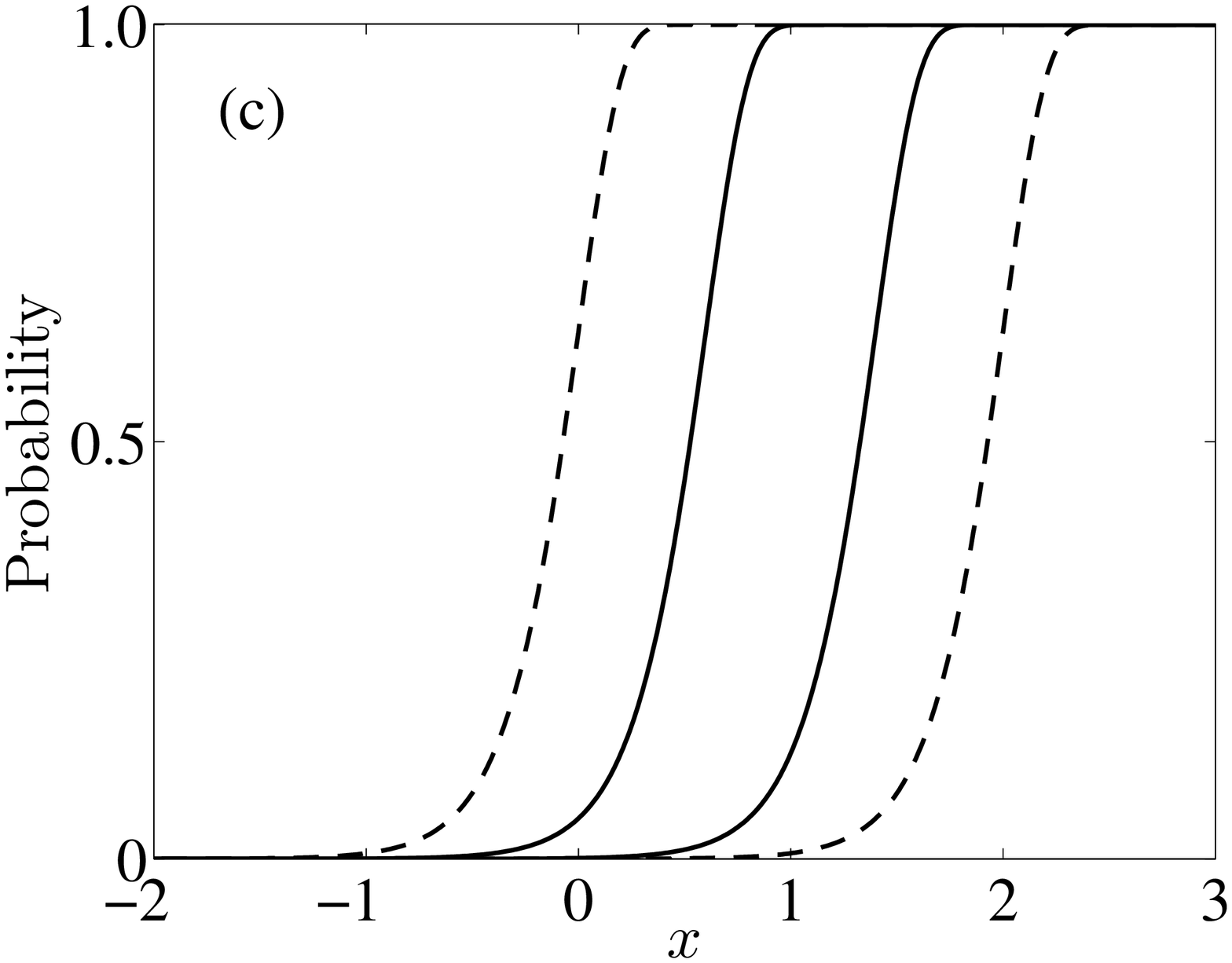}
\caption{The S curves (i.e.~switching probabilities as functions
of readout bias current with fixed pulse duration) for three
different types of detectors: a strongly coupled detector [(a);
see Sec.~IV.B], a weakly coupled fast-decohering detector [(b);
see Sec.~IV.C] and a weakly coupled slow-decohering detector [(c);
see Sec.V]. The dashed lines are the S curves for the qubit's
clockwise- and counterclockwise-current states, i.e.~the ones that
are obtained when the qubit is biased far from the degeneracy
point. In the example plotted in this figure, the probabilities
for these S curves (i.e.~the dashed lines) are given by
$P=1-e^{-\gamma_{L/R} t}$ with $\gamma_L t= e^{5(x-2)}$ and
$\gamma_R t= e^{5x}$ ($x$ is a dimensionless parameter that
represents the bias current used in the readout pulse, and
$\gamma_{L/R}$ represent the switching rates for the two different
persistent-current states). The solid lines are the S curves for
the qubit's energy eigenstates when these states have the form
$\ket{0}=\sqrt{0.7} \ket{L} + \sqrt{0.3} \ket{R}$ and
$\ket{1}=\sqrt{0.3} \ket{L} - \sqrt{0.7} \ket{R}$. In (a) the
switching probabilities for the states $\ket{L}$ and $\ket{R}$ are
averaged (in a weighted manner) in order to obtain the switching
probabilities for the states $\ket{0}$ and $\ket{1}$. This
reasoning is the one that would be consistent with the picture of
a strong measurement. In (b) the switching rates are averaged, and
the switching probabilities are calculated from the averaged
switching rates (see Sec.~IV.C). In (c) the parameter $x$ is
averaged, and from the two weighted averages of $x$ the switching
rates and probabilities are calculated (see Sec.~V). Here we are
treating $x$ as the physical observable that is modified by the
qubit at the Hamiltonian level.}
\end{figure}

\begin{figure}[h]
\includegraphics[width=6.0cm]{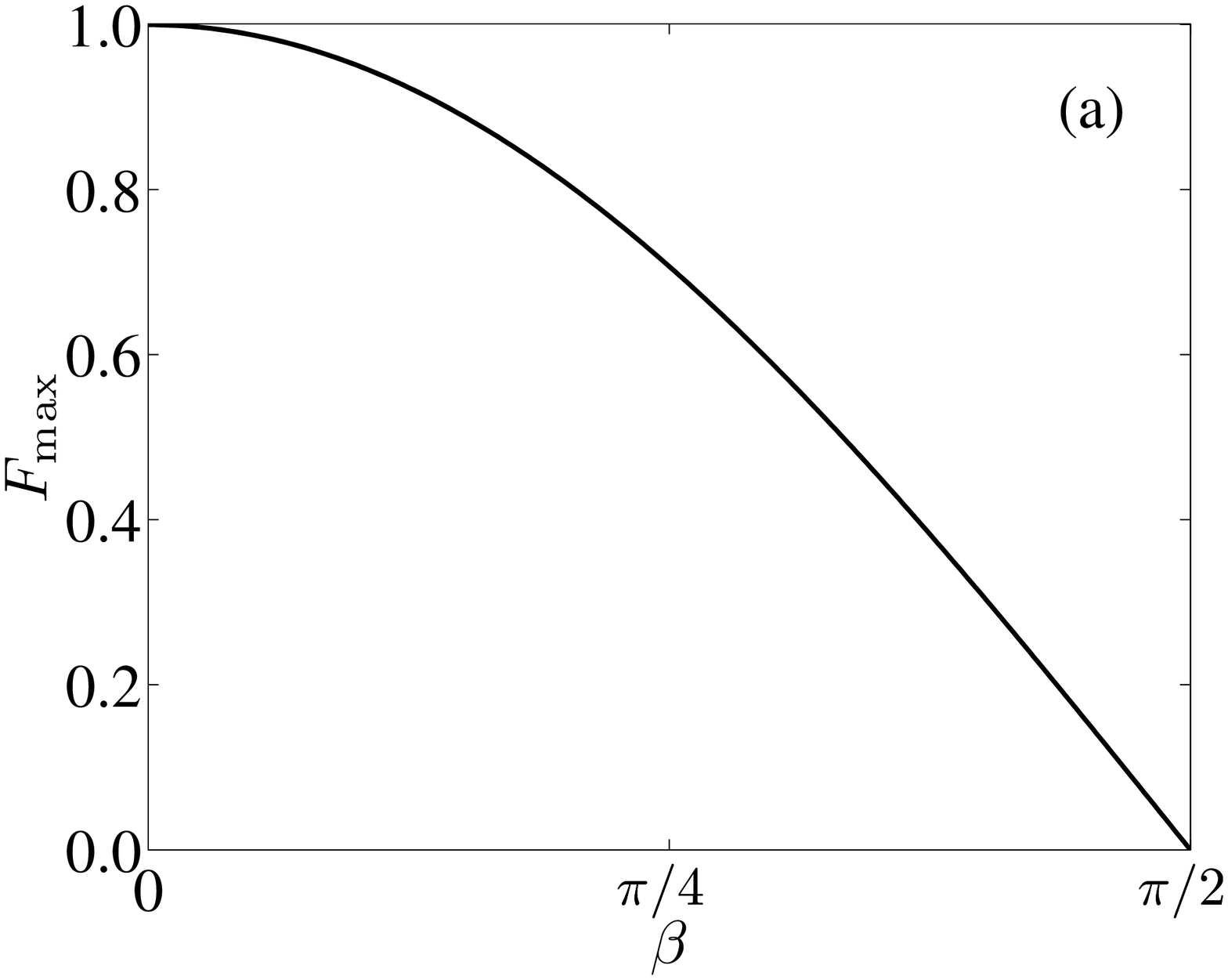}
\includegraphics[width=6.0cm]{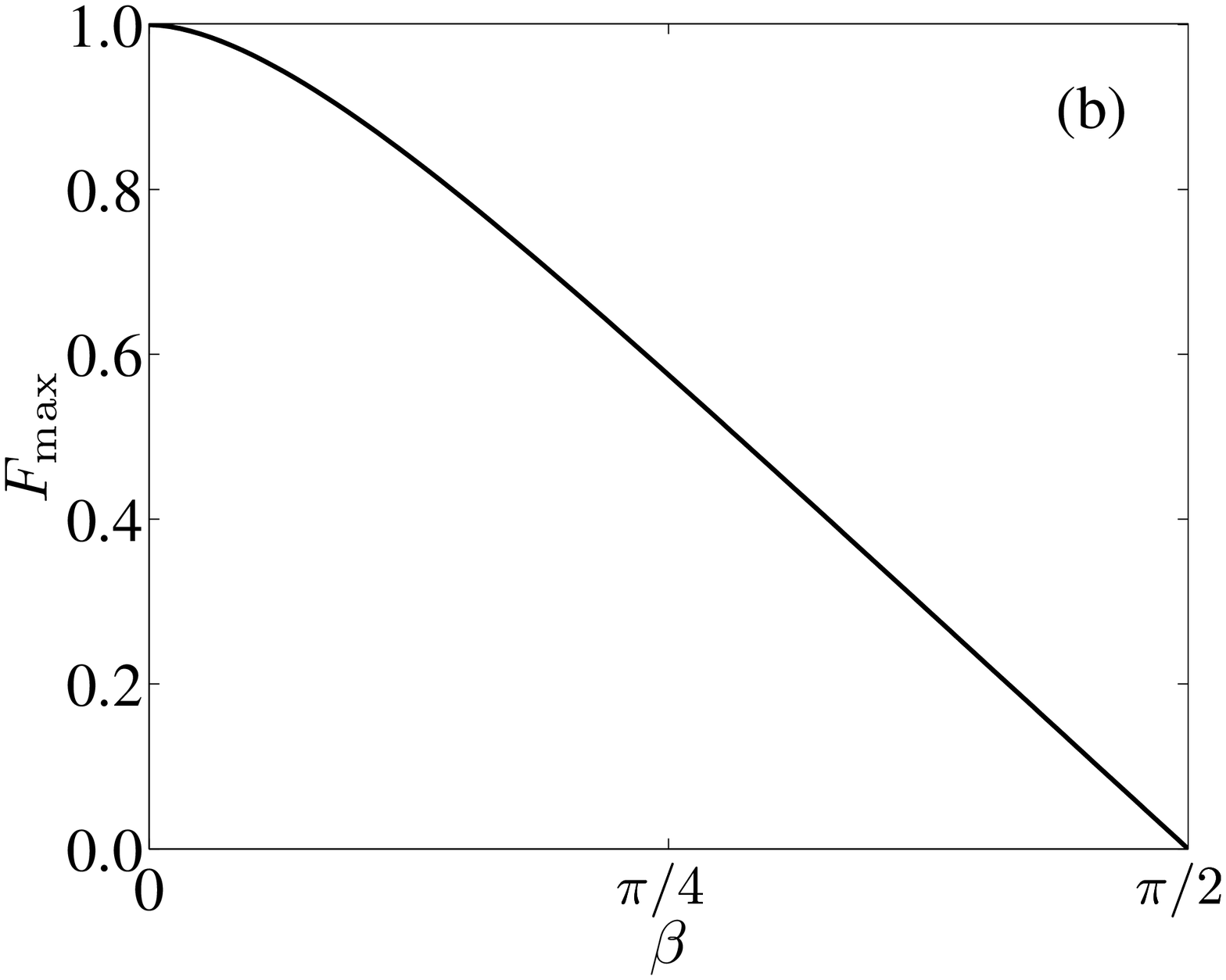}
\includegraphics[width=6.0cm]{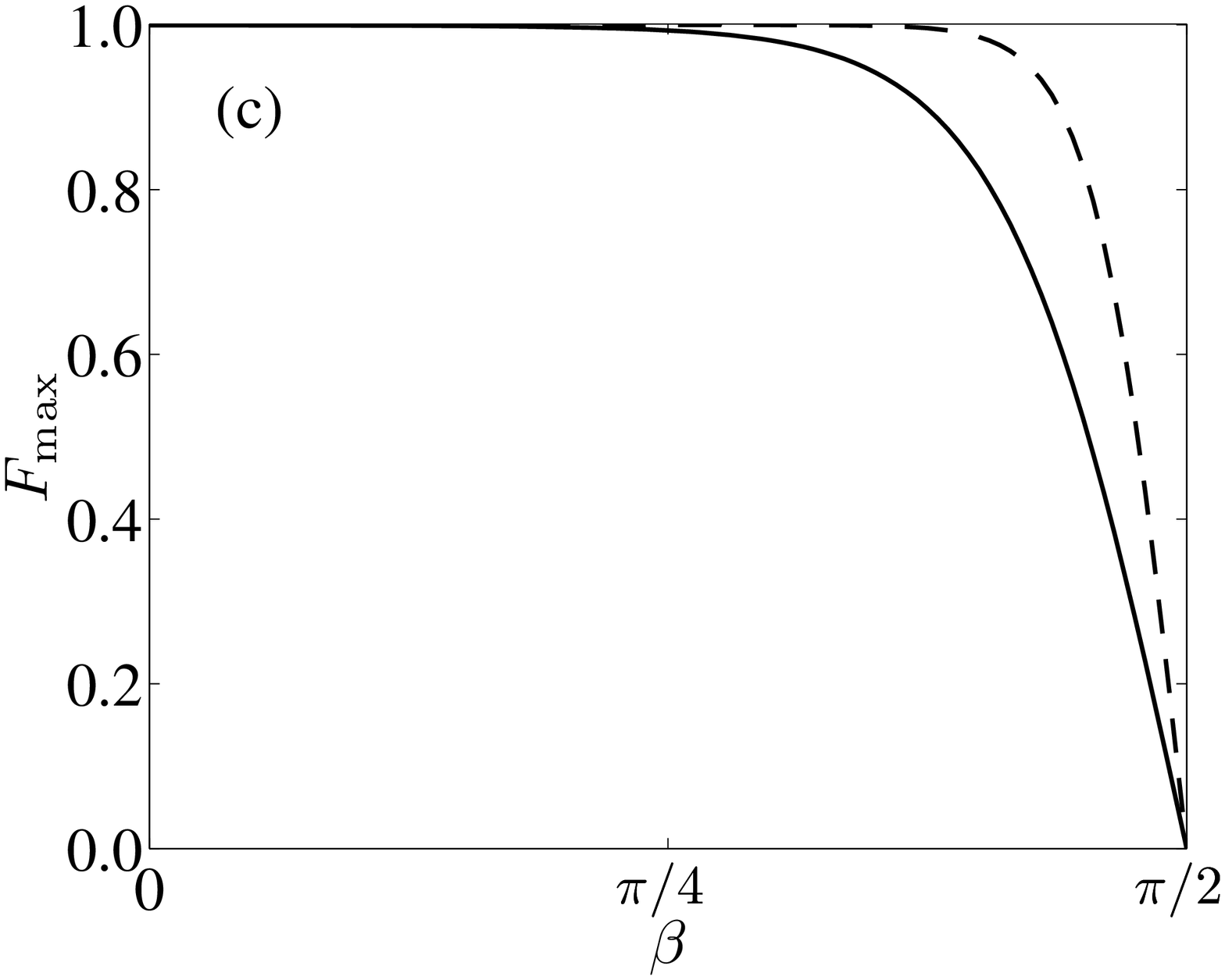}
\caption{The maximum measurement fidelity (i.e.~at the optimal
value of $x$) as a function of the angle $\beta$ (between the
qubit Hamiltonian's axis and the probed operator axis) for the
three different types of detectors whose S curves are plotted in
Fig.~5. The fidelity is derived by finding the maximum vertical
separation between the two S-curves for a given value of $\beta$.
In (a) the fidelity is given by $\cos\beta$, as expected for the
textbook-type projective measurement. In (b) the fidelity is given
by Eq.~(\ref{Eq:Fidelity_incoherent}). In (c) the fidelity is
calculated numerically. For further demonstration that the
coherent detector's fidelity can remain close to 100\% even for
$\beta\neq 0$, we plot in Fig.~6c (dashed line) the fidelity when
the coefficient 5 in the switching-rate formulae of Fig.~5 is
replaced by the coefficient 10 (i.e.~in going from the solid to
the dashed line, the sensitivity of the detector is increased).}
\end{figure}

We now ask the question of what the S curves would look like if
the qubit were biased close to the so-called degeneracy point,
where the energy eigenstates are superpositions of the two
persistent-current eigenstates. Figure 5 shows three possibilities
that correspond to three different types of detectors: a strongly
coupled detector (see Sec.~IV.B), a weakly coupled fast-decohering
detector (see Sec.~IV.C) and a weakly coupled slow-decohering
detector (see Sec.V) [the explanation of the reasoning behind each
one the obtained shapes of the S-curves is given in Sec.~VI]. In
Fig.~6 we plot the measurement fidelity (at the optimal bias
current) for the three different types of detectors. Clearly,
these three types of detectors operate drastically differently.
The fidelity in Fig.~6(a) is given by $\cos\beta$, and the one in
Fig.~6(b) is given by Eq.~(\ref{Eq:Fidelity_incoherent}), i.e.~it
is identical to the one shown in Fig.~4. As mentioned above (see
also \cite{Ashhab}), the measurement fidelity of a coherent
detector (Fig.~6c) can remain close to 100\% even when the qubit
Hamiltonian and the probed operator do not commute.

The operation of the detector when the qubit is biased close to
the degeneracy point can be used to identify whether it performs a
strong- or weak-coupling measurement and how the detector's
coherence time compares to the qubit's precession period. By
tuning the qubit's energy splitting and probing it using the
detector, it could be possible to measure the coherence time of
the detector. This approach could also be used to probe coherent
behaviour in the detector.

Finally, we comment on the description of a continuous-variable
switching detector, which should be a good description for the
dc-SQUID-based measurement of the flux qubit (see also
\cite{Nakano}). Instead of our two-state-detector model, one can
in general use a model of a fictitious particle initialized in a
local minimum of a one-dimensional potential. The clockwise- and
counterclockwise-current states of the qubit correspond to
slightly different potentials for this particle. The two
potentials can result in very different escape rates (out of the
local minimum). We expect that in the case of an incoherent
detector the above results for a two-state detector also hold for
the continuous-variable detector. In the case of a coherent
detector, one can follow the general procedure that we used for
the two-state detector, modified for the continuous-variable
detector. One takes the two trapping potentials for the clockwise-
and counterclockwise-current states. From these potentials one
derives two effective potentials (that correspond to the qubit's
two energy eigenstates) for the fictitious particle, each
potential obtained as a weighted average of the two bare
potentials. The proper proportions of the original potentials that
are used to evaluate the energy-eigenstate effective potentials
are determined by the mixing probabilities of the
persistent-current states in the energy eigenstates. These
weighted-average potentials now determine the relevant escape
rates for the particle, and these escape rates characterize the
operation of the detector. As in Sec.~V (two-state coherent
detector), the detector performs a measurement on the qubit in the
energy eigenbasis. However, as can be seen in Fig.~6(c), the
continuous-variable detector has the advantage that it could lead
to a high fidelity measurement even when the energy eigenstates
are superpositions of persistent-current states. In fact, such
high-fidelity readout in the energy eigenbasis was observed in the
experiment of Ref.~\cite{Tanaka}.

A note is in order here on the relevant coherence time that
separates the coherent-detector and incoherent-detector regimes
for a continuous-variable detector. Even if the coherence time
between the switched (or escaped) state and the non-switched (or
trapped) state is short, the coherence time that corresponds to
motion around the local minimum of the trapping potential could
still be long. If (1) the temperature is low enough such that the
detector's fictitious particle is almost in the ground state of
its effective potential and (2) the coherence time for
superpositions of the ground states of the two possible effective
potentials is longer than $1/E$, the detector would operate in the
coherent regime [Fig.~6(c)]. A full understanding of the boundary
at which decoherence in the detector causes irreversible collapse
of the state of the system and how that affects the state of the
qubit is lacking. Our results demonstrate that the coherence
properties of the detector are crucial for determining its regime
of operation. However, they do not answer questions about the
coherence within the trapped state, or states, of the detector.

\section{Conclusion}

We have analyzed the operation of a switching-based detector that
is used to measure the state of a qubit in the case when the
probed observable does not commute with the qubit Hamiltonian. We
have found several operation regimes depending on the relation
between the detector's switching rate and the qubit's energy
(strong versus weak-coupling measurement) and the relation between
the detector's coherence time and the qubit's energy (coherent
versus incoherent detector). The accessibility of the exact
switching time also plays a role in determining the performed
measurement. Table I summarizes the differences between the
different regimes.

The weak-coupling regimes result in a number of interesting
phenomena. For example, an incoherent detector can be used for an
`all-in-one' measurement setting that can be used to
simultaneously calibrate the system and perform quantum-state
tomography. Apart from isolated special cases, a coherent detector
always measures the qubit in its energy eigenbasis, regardless of
the quantity that it naturally probes. Furthermore, the
measurement fidelity of a coherent detector can remain close to
100\%, even when the qubit's Hamiltonian and the qubit operator
being probed by the detector do not commute.

Although we have analyzed a simplified two-state-detector model
that was inspired by the measurement of superconducting flux
qubits using dc SQUIDs, the main results should apply to other
types of qubits and measurement devices, e.g.~charge qubits
measured using a quantum point contact \cite{Gurvitz}. As far as
the roles of the different coherence and coupling parameters are
concerned, our results provide insight into the operation of a
general type of detector and can be useful for future studies
aimed at designing high-fidelity measurement devices.

We would like to thank P. de Groot, C. Harmans, K. Harrabi, A.
Kofman, A. J. Leggett, Y. Nazarov, T. Picot and H. Wei for useful
discussions, and especially P. Bertet for drawing our attention to
the significance of the S-curves as a diagnostic tool for
characterizing the detector's operation. This work was supported
in part by the National Security Agency (NSA), the Laboratory for
Physical Sciences (LPS), the Army Research Office (ARO) and the
National Science Foundation (NSF) grant No.~EIA-0130383. J.Q.Y.
was also supported by the National Basic Research Program of China
grant No. 2009CB929300, the National Natural Science Foundation of
China grant No. 10625416, and the MOST International Collaboration
Program grant No. 2008DFA01930.

\begin{center}
{\bf Appendix A: \\ Incoherent detector with $E \gg \gamma_-$}
\end{center}

When $E \gg \gamma_-$, Eq.~(\ref{Eq:lambda_and_phi}) gives
\begin{eqnarray}
\lambda_{\pm} & \approx & - \frac{\gamma_+}{2} \pm \frac{1}{2}
\left( \gamma_- \cos\beta + i \left[ E - \frac{\gamma_-^2}{2E}
\right] \right) \nonumber
\\
\cos\frac{\eta}{2} & \approx & 1 \nonumber
\\
\sin\frac{\eta}{2} & \approx & 0,
\end{eqnarray}
which, when substituted into Eq.~(\ref{Eq:U_no_switch}), gives
\begin{equation}
\hat{U}_{\rm ns} (t) = \left(
\begin{array}{cc}
e^{-\frac{\gamma_+}{2} t + \frac{\gamma_-\cos\beta}{2} t + i
\frac{\tilde{E}}{2} t} & 0 \\
0 & e^{-\frac{\gamma_+}{2} t - \frac{\gamma_-\cos\beta}{2} t - i
\frac{\tilde{E}}{2} t}
\end{array}
\right),
\end{equation}
where we have defined $\tilde{E}=E-\gamma_-^2/2E$.

Using Eq.~(\ref{Eq:U_switch}) we find that
\begin{widetext}
\begin{eqnarray}
\hat{U}_{\rm s} (t,\Delta t) & = & e^{-\frac{\gamma_+}{2} t}
\sqrt{\Delta t} \left[ \sqrt{\gamma_L} \left(
\begin{array}{cc}
\cos^2\frac{\beta}{2} & \sin\frac{\beta}{2}\cos\frac{\beta}{2} \\
\sin\frac{\beta}{2}\cos\frac{\beta}{2} & \sin^2\frac{\beta}{2} \\
\end{array}
\right)+ \sqrt{\gamma_R} \left( \begin{array}{cc}
\sin^2\frac{\beta}{2} & -\sin\frac{\beta}{2}\cos\frac{\beta}{2} \\
-\sin\frac{\beta}{2}\cos\frac{\beta}{2} & \cos^2\frac{\beta}{2} \\
\end{array}
\right) \right] \nonumber
\\
& & \hspace{6cm} \times \left(
\begin{array}{cc}
e^{\frac{\gamma_-\cos\beta}{2} t + i \frac{\tilde{E}}{2} t} & 0 \\
0 & e^{- \frac{\gamma_-\cos\beta}{2} t - i \frac{\tilde{E}}{2} t}
\end{array}
\right) \nonumber
\\
& = & e^{-\frac{\gamma_+}{2} t} \sqrt{\Delta t} \left(
\begin{array}{cc}
e^{ i \frac{\tilde{E}}{2} t} & 0 \\
0 & e^{- i \frac{\tilde{E}}{2} t}
\end{array}
\right) \times \nonumber
\\
& & \left[ \sqrt{\gamma_L} \left(
\begin{array}{cc}
e^{\frac{\gamma_-\cos\beta}{2} t} \cos^2\frac{\beta}{2} & e^{-
\frac{\gamma_-\cos\beta}{2} t} e^{- i \tilde{E} t}
\sin\frac{\beta}{2} \cos\frac{\beta}{2} \\
e^{\frac{\gamma_-\cos\beta}{2} t} e^{i \tilde{E} t}
\sin\frac{\beta}{2}\cos\frac{\beta}{2} &
e^{- \frac{\gamma_-\cos\beta}{2} t} \sin^2\frac{\beta}{2} \\
\end{array}
\right) \right. \nonumber \\ & & \left. + \sqrt{\gamma_R} \left(
\begin{array}{cc} e^{\frac{\gamma_-\cos\beta}{2} t}
\sin^2\frac{\beta}{2} & - e^{- \frac{\gamma_-\cos\beta}{2} t}
e^{- i \tilde{E} t} \sin\frac{\beta}{2} \cos\frac{\beta}{2} \\
- e^{\frac{\gamma_-\cos\beta}{2} t} e^{i \tilde{E} t}
\sin\frac{\beta}{2}\cos\frac{\beta}{2} &
e^{- \frac{\gamma_-\cos\beta}{2} t} \cos^2\frac{\beta}{2} \\
\end{array}
\right) \right]. \nonumber
\end{eqnarray}

The above expression gives
\begin{eqnarray}
& & \hspace{-1.5cm} \hat{U}_{\rm s}^{\dagger} (t,\Delta t)
\hat{U}_{\rm s} (t,\Delta t) = e^{-\gamma_+ t} \Delta t \nonumber
\\
& & \times \Bigg[ \sqrt{\gamma_L} \left(
\begin{array}{cc}
e^{\frac{\gamma_-\cos\beta}{2} t} \cos^2\frac{\beta}{2} &
e^{\frac{\gamma_-\cos\beta}{2} t} e^{- i \tilde{E} t}
\sin\frac{\beta}{2} \cos\frac{\beta}{2} \\
e^{-\frac{\gamma_-\cos\beta}{2} t} e^{i \tilde{E} t}
\sin\frac{\beta}{2}\cos\frac{\beta}{2} &
e^{-\frac{\gamma_-\cos\beta}{2} t} \sin^2\frac{\beta}{2} \\
\end{array}
\right) + \nonumber \\ & & \hspace{4cm} \sqrt{\gamma_R} \left(
\begin{array}{cc} e^{\frac{\gamma_-\cos\beta}{2} t}
\sin^2\frac{\beta}{2} & - e^{\frac{\gamma_-\cos\beta}{2} t} e^{- i
\tilde{E} t}
\sin\frac{\beta}{2} \cos\frac{\beta}{2} \\
- e^{-\frac{\gamma_-\cos\beta}{2} t} e^{i \tilde{E} t}
\sin\frac{\beta}{2}\cos\frac{\beta}{2} &
e^{-\frac{\gamma_-\cos\beta}{2} t} \cos^2\frac{\beta}{2} \\
\end{array}
\right) \Bigg] \nonumber
\\
& & \times \Bigg[ \sqrt{\gamma_L} \left(
\begin{array}{cc}
e^{\frac{\gamma_-\cos\beta}{2} t} \cos^2\frac{\beta}{2} &
e^{-\frac{\gamma_-\cos\beta}{2} t} e^{- i \tilde{E} t}
\sin\frac{\beta}{2} \cos\frac{\beta}{2} \\
e^{\frac{\gamma_-\cos\beta}{2} t} e^{i \tilde{E} t}
\sin\frac{\beta}{2}\cos\frac{\beta}{2} &
e^{-\frac{\gamma_-\cos\beta}{2} t} \sin^2\frac{\beta}{2} \\
\end{array}
\right) + \nonumber \\ & & \hspace{4cm} \sqrt{\gamma_R} \left(
\begin{array}{cc} e^{\frac{\gamma_-\cos\beta}{2} t}
\sin^2\frac{\beta}{2} & - e^{-\frac{\gamma_-\cos\beta}{2} t} e^{-
i \tilde{E} t}
\sin\frac{\beta}{2} \cos\frac{\beta}{2} \\
- e^{\frac{\gamma_-\cos\beta}{2} t} e^{i \tilde{E} t}
\sin\frac{\beta}{2}\cos\frac{\beta}{2} &
e^{-\frac{\gamma_-\cos\beta}{2} t} \cos^2\frac{\beta}{2} \\
\end{array}
\right) \Bigg] \nonumber
\\
& & \hspace{-0.5cm} = e^{-\gamma_+ t} \Delta t \nonumber
\\
& & \hspace{0.2cm} \left[ \gamma_L \left(
\begin{array}{cc}
e^{\gamma_-t \cos\beta} \cos^2\frac{\beta}{2} & e^{- i \tilde{E}
t} \sin\frac{\beta}{2} \cos\frac{\beta}{2} \\
e^{i \tilde{E} t} \sin\frac{\beta}{2}\cos\frac{\beta}{2} &
e^{-\gamma_-t \cos\beta} \sin^2\frac{\beta}{2} \\
\end{array}
\right) + \gamma_R \left(
\begin{array}{cc}
e^{\gamma_-t \cos\beta} \sin^2\frac{\beta}{2} & - e^{- i
\tilde{E} t} \sin\frac{\beta}{2} \cos\frac{\beta}{2} \\
- e^{i \tilde{E} t} \sin\frac{\beta}{2}\cos\frac{\beta}{2} &
e^{-\gamma_-t \cos\beta} \cos^2\frac{\beta}{2} \\
\end{array}
\right) \right].
\label{Eq:UdaggerU_for_switching}
\end{eqnarray}

In order to determine the measurement basis for instances where a
switching event occurs between $t$ and $t+\Delta t$, the matrix
$\hat{U}_{\rm s}^{\dagger} (t,\Delta t) \hat{U}_{\rm s} (t,\Delta
t)$ must be diagonalized, i.e.~converted into the form
\begin{equation}
P_1 \left(
\begin{array}{cc}
\cos^2\frac{\theta}{2} & e^{-i \phi} \sin\frac{\theta}{2}
\cos\frac{\theta}{2} \\
e^{i \phi} \sin\frac{\theta}{2}\cos\frac{\theta}{2} &
\sin^2\frac{\theta}{2} \\
\end{array}
\right) + P_2 \left(
\begin{array}{cc}
\sin^2\frac{\theta}{2} & - e^{-i \phi} \sin\frac{\theta}{2}
\cos\frac{\theta}{2} \\
- e^{i \phi} \sin\frac{\theta}{2}\cos\frac{\theta}{2} &
\cos^2\frac{\theta}{2} \\
\end{array}
\right).
\end{equation}
Clearly $\phi=\tilde{E} t$, such that the measurement basis
rotates about the $z$ axis with frequency $\tilde{E}$, similarly
to the free evolution of qubit states but with a slightly reduced
rate (note also that the sense of rotation is opposite to that of
state precession). With straightforward algebra we find that
\begin{eqnarray}
\tan \theta & = & \frac{(\gamma_L-\gamma_R) \sin\beta}{\gamma_L
\left( e^{\gamma_-t \cos\beta} \cos^2\frac{\beta}{2} -
e^{-\gamma_-t \cos\beta} \sin^2\frac{\beta}{2} \right) + \gamma_R
\left( e^{\gamma_-t \cos\beta} \sin^2\frac{\beta}{2} -
e^{-\gamma_-t \cos\beta} \cos^2\frac{\beta}{2} \right)} \nonumber
\\
P_1 + P_2 & = & e^{-\gamma_+ t} \Delta t \left\{ \gamma_L \left(
e^{\gamma_-t \cos\beta} \cos^2\frac{\beta}{2} + e^{-\gamma_-t
\cos\beta} \sin^2\frac{\beta}{2} \right) + \gamma_R \left(
e^{\gamma_-t \cos\beta} \sin^2\frac{\beta}{2} + e^{-\gamma_-t
\cos\beta} \cos^2\frac{\beta}{2} \right) \right\} \nonumber
\\
P_1 - P_2 & = & e^{-\gamma_+ t} \Delta t \frac{(\gamma_L-\gamma_R)
\sin\beta}{\sin\theta}
\end{eqnarray}
\end{widetext}

\begin{center}
{\bf Appendix B: \\ Incoherent detector with $E \gg \gamma_+$ and
$\beta=\pi/2$}
\end{center}

For further demonstration purposes, we analyze in this Appendix
the case of an incoherent detector with $E\gg\gamma_+$ and
$\beta=\pi/2$. In this case Eq.~(\ref{Eq:lambda_and_phi}) reduces
to
\begin{eqnarray}
\lambda_{\pm} & = & - \frac{\gamma_+}{2} \pm \frac{i}{2} \sqrt{E^2
- \gamma_-^2}.
\end{eqnarray}
Equation (\ref{Eq:U_no_switch}) therefore gives
\begin{widetext}
\begin{eqnarray}
\hat{U}_{\rm ns} (t) & = & \left(
\begin{array}{cc}
e^{-\frac{\gamma_+}{2} t + i \frac{\tilde{E}}{2} t} & 0 \\
0 & e^{-\frac{\gamma_+}{2} t - i \frac{\tilde{E}}{2} t}
\end{array}
\right) \nonumber
\\
\hat{U}_{\rm s} (t,\Delta t) & = & \frac{e^{-\frac{\gamma_+}{2} t}
\sqrt{\Delta t}}{2} \left(
\begin{array}{ccc}
e^{i \frac{\tilde{E}}{2} t} \left( \sqrt{\gamma_L} +
\sqrt{\gamma_R} \right) & & e^{- i \frac{\tilde{E}}{2} t} \left(
\sqrt{\gamma_L} - \sqrt{\gamma_R} \right) \\
e^{i \frac{\tilde{E}}{2} t} \left( \sqrt{\gamma_L} -
\sqrt{\gamma_R} \right) & & e^{- i \frac{\tilde{E}}{2} t} \left(
\sqrt{\gamma_L} + \sqrt{\gamma_R} \right) \\
\end{array}
\right),
\end{eqnarray}
\end{widetext}
where $\tilde{E}=\sqrt{E^2 - \gamma_-^2}$. The above two matrices
can now be re-expressed as
\begin{eqnarray}
\hat{U}_{\rm ns} (t) = \left(
\begin{array}{cc}
e^{i \frac{\tilde{E}}{2} t} & 0 \\
0 & e^{- i \frac{\tilde{E}}{2} t}
\end{array}
\right) \hat{U}_{\rm ns}^{\rm meas}(t) \nonumber
\\
\hat{U}_{\rm s} (t,\Delta t) = \left(
\begin{array}{cc}
e^{i \frac{\tilde{E}}{2} t} & 0 \\
0 & e^{- i \frac{\tilde{E}}{2} t} \\
\end{array}
\right) \hat{U}_{\rm s}^{\rm meas} (t,\Delta t),
\end{eqnarray}
where
\begin{eqnarray}
\hat{U}_{\rm ns}^{\rm meas}(t) & = & \left(
\begin{array}{cc}
e^{-\frac{\lambda_+}{2} t} & 0 \\
0 & e^{-\frac{\lambda_+}{2} t}
\end{array}
\right) \nonumber
\\
\hat{U}_{\rm s}^{\rm meas} (t,\Delta t) & = &
e^{-\frac{\gamma_+}{2} t} \left\{ \frac{\sqrt{\gamma_L \Delta
t}}{2} \left(
\begin{array}{cc}
1 & e^{- i \tilde{E} t} \\
e^{i \tilde{E} t} & 1 \\
\end{array}
\right) + \right. \nonumber
\\
& & \hspace{1.5cm} \left. \frac{\sqrt{\gamma_R \Delta t}}{2}
\left(
\begin{array}{cc}
1 & - e^{- i \tilde{E} t} \\
- e^{i \tilde{E} t} & 1 \\
\end{array}
\right) \right\} \nonumber
\\
& = & e^{-\frac{\gamma_+}{2} t} \left( \sqrt{\gamma_L \Delta t}
\ket{\psi_{\rm s}^{(1)}(t)} \bra{\psi_{\rm s}^{(1)}(t)} + \right.
\nonumber
\\
& & \hspace{1.5cm} \left. \sqrt{\gamma_R \Delta t} \ket{\psi_{\rm
s}^{(2)}(t)} \bra{\psi_{\rm s}^{(2)}(t)} \right) \nonumber
\end{eqnarray}
\begin{eqnarray}
\ket{\psi_{\rm s}^{(1)}(t)} & = & \frac{1}{\sqrt{2}} \left(
\begin{array}{c}
e^{- i \frac{\tilde{E}}{2} t} \\
e^{i \frac{\tilde{E}}{2} t} \\
\end{array}
\right) = \left(
\begin{array}{cc}
e^{-i \frac{\tilde{E}}{2} t} & 0 \\
0 & e^{i \frac{\tilde{E}}{2} t} \\
\end{array}
\right) \left(
\begin{array}{c}
\frac{1}{\sqrt{2}} \\
\frac{1}{\sqrt{2}} \\
\end{array}
\right) \nonumber
\\
\ket{\psi_{\rm s}^{(2)}(t)} & = & \frac{1}{\sqrt{2}} \left(
\begin{array}{c}
e^{- i \frac{\tilde{E}}{2} t} \\
- e^{i \frac{\tilde{E}}{2} t} \\
\end{array}
\right) = \left(
\begin{array}{cc}
e^{-i \frac{\tilde{E}}{2} t} & 0 \\
0 & e^{i \frac{\tilde{E}}{2} t} \\
\end{array}
\right) \left(
\begin{array}{c}
\frac{1}{\sqrt{2}} \\
-\frac{1}{\sqrt{2}} \\
\end{array}
\right).
\nonumber \\
\end{eqnarray}

The interpretation of these results is obvious. In the case of no
switching, the qubit state evolves freely according to the
Hamiltonian (i.e.~rotates around the $z$ axis by an angle
approximately given by $Et$), but we essentially do not gain any
information about the state of the qubit. On the other hand, an
experimental run with the outcome of no switching until time $t$
followed by a switching event between times $t$ and $t+\Delta t$
is equivalent to a measurement in the basis $\left\{
\ket{\psi_{\rm s}^{(1)}(t)}, \ket{\psi_{\rm s}^{(2)}(t)} \right\}$
followed by a rotation around the $z$ axis by an angle
$\tilde{E}t$. The fidelity of the measurement is given by
\begin{equation}
F_{\rm s}(t,\Delta t) = \frac{\gamma_R - \gamma_L}{\gamma_R +
\gamma_L}.
\end{equation}
It is interesting that the fidelity is constant in time. The only
change that occurs in time is that the measurement is performed
along a rotated axis that precesses about the axis of the qubit
Hamiltonian at rate $\tilde E$. As explained above, the
measurement axis is determined only at the time that the detector
switches. Of course an equivalent description of the measurement
dynamics would be to say that the qubit state precesses freely
around the $z$ axis by an angle $\tilde{E}t$ and is then measured
in the $\left\{ \ket{L}, \ket{R} \right\}$ basis when the detector
switches. Depending on whether one is interested in reaching an
intuitive description of the qubit-state evolution or in
identifying what information can be extracted regarding the
initial state of the qubit, one or the other of the above
descriptions would be more relevant.

We can now use the above results to demonstrate again the
advantage of keeping track of the exact switching time. If the
switching time is not known (up to the pulse duration $\tau$,
which we take to be much longer than $1/E$), one must use the
matrices $\hat{U}_{\rm ns}^{\dagger}(\tau) \hat{U}_{\rm ns}(\tau)$
and $\hat{U}_{\rm s}^{\dagger}(0,\tau) \hat{U}_{\rm s}(0,\tau)$ as
the measurement matrices. The two eigenvalues of each one of the
above matrices are equal. Therefore, the fidelity vanishes. This
result is in agreement with the generally accepted rule that this
type of measurement cannot be used when the qubit is biased at the
so-called degeneracy point during the measurement. This example
shows clearly that additional information can be obtained by
keeping track of the exact switching time, provided this switching
time can be determined with accuracy smaller than $1/E$.


\begin{thebibliography}{99}
\bibitem{Single_setting} The term `single setting' should not be
confused with the term `single shot'. One should note here that
quantum state tomography cannot be performed in a single
experimental run. Averaging of the results over a large number of
identically prepared states is necessary for full state
tomography.
\end{thebibliography}
\end{document}